\documentclass[apj]{emulateapj}
\newcommand{\arone}{148\,GHz}
\newcommand{\artwo}{218\,GHz}
\newcommand{\arthree}{277\,GHz}

\newcommand{\commentx}[1]{}

\renewcommand{\vec}[1]{\mbox{\boldmath$#1$}} % Bold
\newcommand{\ra}[3]   % right ascension
   {\makebox[1.5em][r]{#1}\makebox[1.5em][r]{#2} \makebox[2em][r]{#3}}

\newcommand{\hms}[3]  % Write HMS in ApJ style.
   {${#1}^{\mathrm{h}}{#2}^{\mathrm{m}}{#3}^{\mathrm{s}}$}

\newcommand{\hmin}[2]  % Write HM in ApJ style.
   {\ensuremath{{#1}^{\mathrm{h}}{#2}^{\mathrm{m}}}}
   
\newcommand{\hours}[1]  % Write H in ApJ style.
   {\ensuremath{{#1}^{\mathrm{h}}}}

\newcommand{\dms}[3]  % Write DMS in ApJ style.
   {\ensuremath{{#1}\degree{#2}\arcminute{#3}\arcsecond}}

\newcommand{\dm}[2]  % Write DM in ApJ style.
   {\ensuremath{{#1}\degree{#2}\arcminute}}

\newcommand{\ukcmb}  % microKelvin_cmb
           {\ensuremath{\micro \kelvin_\mathrm{cmb}}}

\newcommand{\uk}  % microKelvin
           {\ensuremath{\micro \kelvin}}

\newcommand{\fdeg} % fractional degrees
           {\hbox{$.\!\!^{\circ}$}}

\usepackage{color}

\usepackage[citebordercolor={0 .5 .5}]{hyperref}

\newcommand{\beq}{\begin{equation}}
\newcommand{\eeq}{\end{equation}}
\newcommand{\be}{\begin{equation}}
\newcommand{\ee}{\end{equation}}
\newcommand{\bea}{\begin{eqnarray}}
\newcommand{\eea}{\end{eqnarray}}
\newcommand{\bdi}{\begin{displaymath}}
\newcommand{\edi}{\end{displaymath}}

\newcommand{\rmicron}{$\,\micro$m}%{$\mu$m}
\def\lsim{\,\lower2truept\hbox{${<\atop\hbox{\raise4truept\hbox{$\sim$}}}$}\,}
\def\gsim{\,\lower2truept\hbox{${>\atop\hbox{\raise4truept\hbox{$\sim$}}}$}\,}
\newcommand{\ndivtext}{four }

\usepackage{natbib}  % Use the "Natbib" style for the
\usepackage{subfigure}  % Use the "Natbib" style for the
\usepackage{rotating}
\usepackage{longtable}
\usepackage[mediumspace,Gray,squaren]{SIunits}

\shorttitle{Correlations in the CSB from $\rm ACT\times BLAST$}

\shortauthors{Hajian, Viero, et al.}

\begin{document}

\title{Correlations in the (Sub)millimeter background from $\rm ACT \times BLAST$}

%%%%%%%%%%%%%%%%%%%%%%%%%%%%%%%%%%%%%%%%%%%%%%%%%%%%%%%%%%%%%%%%%%%%%%%%%%%
% WARNING: This LaTeX block was generated automatically by authors.py
% Do not change by hand: your changes will be lost.

\author{
Amir~Hajian\altaffilmark{1,2,3},
Marco~P.~Viero\altaffilmark{4,5},
Graeme~Addison\altaffilmark{6},
Paula~Aguirre\altaffilmark{7},
John~William~Appel\altaffilmark{3},
Nick~Battaglia\altaffilmark{1},
James~J.~Bock\altaffilmark{8,4},
J.~Richard~Bond\altaffilmark{1},
Sudeep~Das\altaffilmark{9,3,2},
Mark~J.~Devlin\altaffilmark{10},
Simon~R.~Dicker\altaffilmark{10},
Joanna~Dunkley\altaffilmark{6,3,2},
Rolando~D\"{u}nner\altaffilmark{3},
Thomas~Essinger-Hileman\altaffilmark{3},
John~P.~Hughes\altaffilmark{24},
Joseph~W.~Fowler\altaffilmark{11,3},
Mark~Halpern\altaffilmark{12},
Matthew~Hasselfield\altaffilmark{12},
Matt~Hilton\altaffilmark{16,17},
Adam~D.~Hincks\altaffilmark{3},
Ren\'ee~Hlozek\altaffilmark{6},
Kent~D.~Irwin\altaffilmark{11},
Jeff~Klein\altaffilmark{10},
Arthur~Kosowsky\altaffilmark{13},
Yen-Ting~Lin\altaffilmark{23},
Tobias~A.~Marriage\altaffilmark{2,14},
Danica~Marsden\altaffilmark{10},
Gaelen~Marsden\altaffilmark{12},
Felipe~Menanteau\altaffilmark{24},
Lorenzo~Moncelsi\altaffilmark{15},
Kavilan~Moodley\altaffilmark{16},
Calvin~B.~Netterfield\altaffilmark{5,18},
Michael~D.~Niemack\altaffilmark{11,3},
Michael~R.~Nolta\altaffilmark{1},
Lyman~A.~Page\altaffilmark{3},
Lucas~Parker\altaffilmark{3},
Douglas~Scott\altaffilmark{12},
Neelima~Sehgal\altaffilmark{19},
Jon~Sievers\altaffilmark{1},
David~N.~Spergel\altaffilmark{2},
Suzanne~T.~Staggs\altaffilmark{3},
Daniel~S.~Swetz\altaffilmark{10,11},
Eric~R.~Switzer\altaffilmark{20,3},
Robert~Thornton\altaffilmark{10,21},
Ed~Wollack\altaffilmark{22},
}
\altaffiltext{1}{Canadian Institute for Theoretical Astrophysics, University of
Toronto, Toronto, ON\ M5S~3H8, Canada}
\altaffiltext{2}{Department of Astrophysical Sciences, Peyton Hall, 
Princeton University, Princeton, NJ\ 08544, USA}
\altaffiltext{3}{Joseph Henry Laboratories of Physics, Jadwin Hall,
Princeton University, Princeton, NJ\ 08544, USA}
\altaffiltext{4}{California Institute of Technology, 1200 E. California Blvd., P
asadena, CA\ 91125, USA}
\altaffiltext{5}{Department of Astronomy \& Astrophysics, University of Toronto, 50 St. George Street, Toronto, ON\ M5S~3H4, Canada}
\altaffiltext{6}{Department of Astrophysics, Oxford University, Oxford, OX1 3RH, UK}
\altaffiltext{7}{Departamento de Astronom{\'{i}}a y Astrof{\'{i}}sica, 
Facultad de F{\'{i}}sica, Pontific\'{i}a Universidad Cat\'{o}lica,
Casilla 306, Santiago 22, Chile}
\altaffiltext{8}{Jet Propulsion Laboratory, Pasadena, CA\ 91109, USA}
\altaffiltext{9}{Berkeley Center for Cosmological Physics, LBL and
Department of Physics, University of California, Berkeley, CA\ 94720, USA}
\altaffiltext{10}{Department of Physics and Astronomy, University of
Pennsylvania, 209 South 33rd Street, Philadelphia, PA\ 19104, USA}
\altaffiltext{11}{NIST Quantum Devices Group, 325
Broadway Mailcode 817.03, Boulder, CO\ 80305, USA}
\altaffiltext{12}{Department of Physics and Astronomy, University of
British Columbia, Vancouver, BC\ V6T~1Z4, Canada}
\altaffiltext{13}{Department of Physics and Astronomy, University of Pittsburgh, 
Pittsburgh, PA\ 15260, USA}
\altaffiltext{14}{Deptartment of Physics and Astronomy, The Johns Hopkins University, 3400 N. Charles St., Baltimore, MD\ 21218, USA}
\altaffiltext{15}{Department of Physics \& Astronomy, Cardiff University, 5 The Parade, Cardiff, CF24 3AA, UK}
\altaffiltext{16}{Astrophysics and Cosmology Research Unit, School of
Mathematical Sciences, University of KwaZulu-Natal, Durban, 4041,
South Africa}
\altaffiltext{17}{School of Physics \& Astronomy, University of Nottingham, NG7 2RD, UK}
\altaffiltext{18}{Department of Physics, University of Toronto, 
60 St. George Street, Toronto, ON\ M5S~1A7, Canada}
\altaffiltext{19}{Kavli Institute for Particle Astrophysics and Cosmology, Stanford University, Stanford, CA\ 94305, USA}
\altaffiltext{20}{Kavli Institute for Cosmological Physics, 
5620 South Ellis Ave., Chicago, IL\ 60637, USA}
\altaffiltext{21}{Department of Physics , West Chester University 
of Pennsylvania, West Chester, PA\ 19383, USA}
\altaffiltext{22}{Code 553/665, NASA/Goddard Space Flight Center,
Greenbelt, MD\ 20771, USA}
\altaffiltext{23}{Institute for the Physics and Mathematics of the Universe, 
The University of Tokyo, Kashiwa, Chiba 277-8568, Japan}
\altaffiltext{24}{Department of Physics and Astronomy, Rutgers, 
The State University of New Jersey, Piscataway, NJ USA 08854-8019}

% End auto-generated block
%%%%%%%%%%%%%%%%%%%%%%%%%%%%%%%%%%%%%%%%%%%%%%%%%%%%%%%%%%%%%%%%%%%%%%%%%%%

\begin{abstract}
We present measurements of the auto- and cross-frequency correlation power spectra of the cosmic (sub)millimeter background at: 250, 350, and 500\rmicron\ (1200, 860, and 600 GHz) from observations made with the Balloon-borne Large Aperture Submillimeter Telescope, BLAST; and at 1380 and 2030 \rmicron\ (218 and 148 GHz) from observations made with the Atacama Cosmology Telescope, ACT.  The overlapping observations cover $ 8.6~\rm deg^2$ in an area relatively free of Galactic dust near the south ecliptic pole (SEP).  The ACT bands are sensitive to radiation from the CMB, the Sunyaev-Zel'dovich (SZ) effect from galaxy clusters, and to emission by radio and dusty star-forming galaxies (DSFGs), while the dominant contribution to the BLAST bands is from DSFGs.  
We confirm and extend the BLAST analysis of clustering with an independent pipeline, %
and also detect correlations between the ACT and BLAST maps at over 25$\sigma$ significance, which we interpret as a detection of the DSFGs in the ACT maps. 
 In addition to a Poisson component in the cross-frequency power spectra, we detect a clustered signal at $\> 4 \sigma$, and using a model for the DSFG evolution and number counts,  we successfully fit all our spectra  with a linear clustering model and a bias that depends only on redshift and not on scale. Finally, the data are compared to, and generally agree with, phenomenological models for the DSFG population. 
This study represents a first of its kind, and demonstrates the constraining power of the cross-frequency correlation technique to constrain models for the DSFGs.  Similar analyses with more data will impose tight constraints on future models.  
\end{abstract}


% ======================
\section{Introduction}
% ======================

Roughly half of all the light in the extragalactic sky which originated from stars appears as a nearly uniform cosmic infrared background \citep[CIB;][]{puget1996, fixsen1998}.  This background peaks in intensity at around 200\rmicron\ \citep{dole2006}, and results from thermal re-radiation of optical and UV starlight by dust grains, meaning that half of all the light emitted by stars is hidden by a veil of dust.

Following its discovery, stacking analyses have statistically resolved most of the CIB shortward of 500\rmicron\ into discrete, dusty star-forming galaxies (DSFG), and to a lesser extent radio galaxies, at $z\le 3$ \citep[e.g.,][]{dole2006, devlin2009, marsden2009, pascale2009}.   
Longward of 500\rmicron, the contribution from radio galaxies and higher-redshift DSFG to the CIB increases dramatically with increasing wavelength \citep[e.g.,][]{bethermin2010}, and as a result, the CIB at these wavelengths has yet to be fully resolved into discrete sources  \citep[e.g.,][]{zemcov2010c}.   
 At wavelengths longward of $\sim 1~\rm mm$, while both radio sources and DSFGs \citep[e.g.,][]{weiss2009,vieira2010} are still present, signal from the cosmic microwave background (CMB) becomes visible and dominates the power on angular scales larger than $\sim 7~\rm arcmin~ (\ell \sim 3000)$ at $\lambda =2$ mm.

To fully realize the cosmological information encoded in the CMB power spectrum, contributions to it from sources must be removed.  At current mm-wave detection and resolution levels, the radio sources are primarily discrete (Poisson) while the DSFGs are confusion limited and clustered.  For example, at 148 GHz the power spectrum of DSFGs roughly equals the CMB power spectrum at $\ell\approx 3000$. Thus knowledge of DSFGs is important for understanding the scalar spectral index of the primordial fluctuations and other parameters encoded in the high--$\ell$ CMB power spectrum.  

Because CMB maps contain signal from multiple contributors, determining precisely the level at which galaxies
contribute to the CMB power spectra is non-trivial.  
Submillimeter (submm) maps, on the other hand, for the most part contain signal from dusty galaxies, so that cross-frequency correlations of submm and mm-wave maps provide a unique way to isolate the contribution of DSFGs to the CMB maps.  However, submm maps of adequate area and depth have until now not existed.

Here we present the first measurement of the cross-frequency power spectra of submm and mm-wave maps.  
We use mm-wave data from the Atacama Cosmology Telescope \citep[ACT;][]{fowler2007,swetz2010} at 1380 and 2030 \rmicron\ (218 and 148 GHz), collected during the 2008 observing season, and submm wave data from the Balloon-borne Large Aperture Submillimeter Telescope \citep[BLAST;][]{pascale2008, devlin2009} at 250, 350, and 500\rmicron\ (1200, 860, and 600 GHz), which were collected during its 11 day flight, at $\sim 40~\rm  km$ altitude, in Antarctica in 2006.
We use these to measure the power from DSFGs, both Poisson and clustered.  
These results will complement those anticipated from the \emph{Planck} mission \citep{tauber2010} by extending to higher resolution in the mm-wave regime.

This paper is organized as follows:  In \S~\ref{sec:smb} we briefly overview the sources of signal in the submm and mm-wave sky, their spectral signatures, and the models we adopt to describe them.  In \S~\ref{sec:data} and \ref{sec:formalism} we describe the data and detail the techniques used to measure the power spectra.  We present our results in \S~\ref{sec:results}, and interpret them in terms of a linear clustering model in \S~\ref{sec:bias}.  %
We discuss and conclude in \S~\ref{sec:discussion} and \ref{sec:conclusion}.

% ======================
% ======================

\section{The (Sub)millimeter Background}
\label{sec:smb}

The dominant contribution to the cosmic submillimeter and millimeter-wave background, referred to hereafter as the CSB, depends strongly on wavelength and angular scale. One map may have contributions from galaxies, CMB, and the Sunyaev-Zel'dovich (SZ) effect simultaneously.

\subsection{Power Spectra}

The beam-corrected power spectrum of the sky is a superposition, depending on wavelength, of the following terms:
\begin{eqnarray}
C_{\ell}^{ \rm sky} &=& C_{\ell}^{\rm cirrus} + C_{\ell}^{\rm CMB}  + C_{\ell}^{\rm radio} \nonumber \\
&+& C_{\ell}^{\rm DSFG} +  C_{\ell}^{\rm SZ}  + C_{\ell}^{\rm ff} + N_{\ell},
\label{eqn:skysignal}
\end{eqnarray}
where $C_{\ell}$ represents the angular power spectrum in multipole space, $\ell$, and $N_{\ell}$ is the noise.  
Here \lq\lq cirrus\rq\rq\ refers to emission from Galactic dust (\S~\ref{sec:cirrus}), \lq\lq ff\rq\rq\ refers to free-free emission, and \lq\lq radio\rq\rq\ refers to radio sources, whose flux increases at
longer wavelengths (\S~\ref{sec:radio}).  It is assumed that diffuse synchrotron emission is negligible.  For the purposes of this paper we consider GHz-peaked sources and similar objects as radio sources.
Equation~\ref{eqn:skysignal} assumes that the various components are uncorrelated,
when in reality, correlations among various components likely exist.  %
Typically, however, these correlations should be small and can be reasonably neglected.  %
Furthermore, we define the power from the extragalactic sky as $C_l^{ \rm CSB} \equiv C_l^{ \rm sky} - C_l^{\rm cirrus} -C_{\ell}^{\rm ff} $. The $C_{\ell}^{\rm ff} $ component is negligible for the area of the sky we are dealing with and we ignore it.
  In what follows we report cross power spectra as both $C_\ell$ and $P(k_\theta)$, with $k_\theta$ the angular wavenumber. To convert from multipole $\ell$ to $k_{\theta}$, or from  $\mu \rm K^2$ to $\rm Jy^2~sr^{-1}$, see Appendix~\ref{sec:convert}.

In order to isolate the spectra of one or more contributors to the background requires removal, or adequate modeling, of the unwanted power.  Since the contributors have distinct spectral signatures (i.e., their flux densities vary from band to band differently), multi-frequency observations make decomposition of the signal possible.  For discrete sources, the ratio of flux densities from band to band is 
\begin{equation}
\frac{S_{\nu_1}}{S_{\nu_2}} = \left( \frac{\nu_1}{\nu_2} \right) ^{\alpha_{\nu_1 - \nu_2}},
\end{equation}
where $\alpha_{\nu_1 - \nu_2}$ is the \lq\lq spectral index\rq\rq, and is a function of the rest-frame spectral energy distributions (SEDs) of the sources that make up the galaxy population, and their redshift distributions.  
Consequently, measurements of the spectral indices can place powerful constraints on source population models \citep[e.g.,][]{marsden2010, bethermin2010}.

% ======================

\subsection{CMB}
\label{sec:cmb}
At wavelengths longer than $\sim 1~\rm mm$  ($\nu = 350$ GHz) the CMB dominates the power spectrum on scales greater than $\sim 8\arcmin $.  Multiple peaks in the spectrum have been measured most recently by: \citet[][]{
brown2009,%brown/etal:2009, friedman/etal:2009,%}, APEX-SZ\citep{
friedman2009,
reichardt2009a,%reichardt/etal:2009a,%}, ACBAR \citep{
reichardt2009b,%reichardt/etal:2009,%}, SZA\citep{
sayers2009,%sayers/etal:2009,%}, QUaD\citep{
lueker2010,%lueker/etal:2010,% hall/etal:prep,%}, and ACT \citep{
sharp2010,%sharp/etal:2010,%}, BIMA \citep{
fowler2010,%fowler/etal:prep,
das2010}; and  %das/etal:prep
\citet{nolta2009}.  %
Secondary anisotropies include: CMB lensing, which acts to smooth out the peaks and add excess power to the damping tail; and the SZ effect, which distorts the primordial CMB signal.

In the present analysis, the CMB power in ACT maps, which dominates on large scales, does not correlate with signal in the BLAST maps; however, it does act to increase the noise on those scales (see Appendix \ref{app:errorbars}).

% ======================
\subsection{Dusty Star-Forming Galaxies}
\label{sec:dsfg}
DSFGs, as their name implies, are galaxies undergoing vigorous star formation, much of which is optically obscured by dust.  They have average flux densities of $5~\rm mJy$ \citep[at 250\rmicron;][]{marsden2009}, star-formation rates (SFRs) of $\sim 100$ -- $200~\rm M_{\odot} yr^{-1}$ \citep{pascale2009, moncelsi2010}, number densities of $\sim 2\times 10^{-4} ~\rm Mpc^{-3}$ \citep[e.g.,][]{vdokkum2009b}, and typically lie at redshifts 0 -- 4, with the peak in the distribution at $z \sim 2$ \citep{amblard2010}.
They are distinguished from \lq\lq submillimeter galaxies\rq\rq\ (SMGs) discovered by SCUBA \citep{smail1997, hughes1998, eales1999}, which are ten times less abundant \citep{coppin2006}, lie at slightly higher redshifts \citep{chapman2005}, have $\rm SFR\sim 1000 ~\rm M_{\odot} yr^{-1}$, and are thought to be triggered largely by major mergers \citep{engel2010}. 
They are of course related: SMGs comprise the extreme, high-redshift end of the DSFG population.

The dust in DSFGs absorbs starlight and re-emits it in the IR/submm, with a spectral energy distribution (SED) phenomenologically well approximated by a modified blackbody,
\begin{equation}
 S_{\nu} \propto \nu^{\beta} B(\nu),
 \label{eqn:bb}
 \end{equation}
where $B(\nu)$ is the Planck function, and $\beta$ is the emissivity index \citep*[whose value typically spans 1.5 -- 2; e.g.,][]{draine1984}.  The SED of a typical DSFG with temperature $\sim 30~\rm K$ and $\beta = 2$ \citep{chapin2010} peaks at rest-frame $\lambda \simeq 100$\rmicron\ (which redshifts into the submillimeter at $z \sim 1-10$).
 A property of this shape is that with increasing redshifts,
observations in the (sub)mm bands continue to sample at a rest-frame
wavelength close to the peak of the SED, so that even though sources
become more distant, their apparent flux remains roughly constant.
This so-called negative K-correction makes observations at longer
wavelengths more sensitive to higher redshift sources  \citep{blain2003}. As a result, sources at $z\gsim 1$ have a significant impact on the power spectrum at  (sub)mm bands.

The cross-power spectrum between frequency bands 1 and 2 arising from unclustered sources is related to the number counts, i.e., the surface density $N$ as
a function of flux density ($S$) as follows
\begin{equation}
\label{eqn:shot_level}
C_{\ell}^{\rm Poisson}=\int^{S_{\rm cut_1}}_0 \int^{S_{\rm cut_2}}_0S_1 S_2 \frac{d^2N}{dS_1 dS_2} dS_1 dS_2,
\end{equation}
where $dN/dS~\Delta S$ is the number of sources per unit solid angle in a flux bin of width $\Delta S$, at frequency bands 1 and 2, and $S_{\rm cut}$ is the flux density at which the counts are truncated.   
When the slope of the counts is steeper than $-3$, the power diverges at low flux densities, and the power spectrum is dominated by the contribution to the background from faint sources.
In the case of DSFGs, the strong evolution of the source counts with redshift results in a steep slope at the faint end \citep{devlin2009},
so that after masking local sources (with $S_{\rm cut} \lsim 500$ mJy) the DSFG component of the CIB at $\lambda > 250$\rmicron\ is dominated by faint sources and remains finite.

Since galaxies are spatially correlated (being biased tracers of the underlying dark matter field), the DSFG power spectrum has both Poisson and clustered components:
\begin{equation}
C_{\ell}^{\rm DSFG}= C_{\ell}^{\rm DSFG, Poisson} + C_{\ell}^{\rm DSFG, clustered}.
\end{equation}
The measured strength of the clustered component is such that it dominates over the Poisson on scales $\gsim 3$\arcmin\ 
% (\ell \lsim 7150)$ 
\citep[][hereafter V09]{viero2009} and \citep{marsden2009, hall2010, dunkley2010, shirokoff2010}, with the exact value depending on frequency and flux cut.  
How the strengths of the Poisson and clustering terms scale with wavelength, and if they  evolve together or independently, remain an open questions.

% ======================
\begin{figure*}[ht]
\centering
\includegraphics[width=0.9\textwidth]{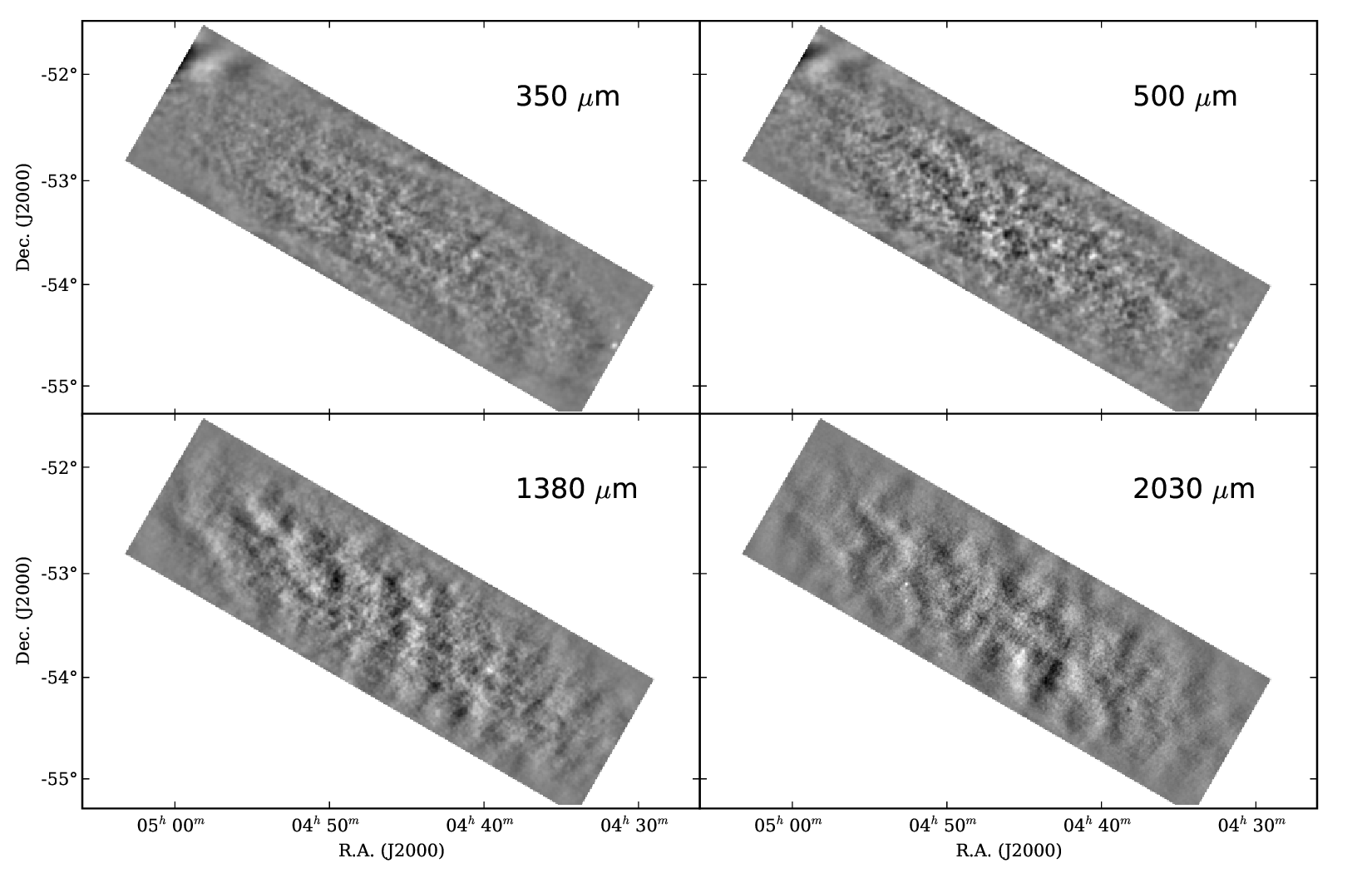}
\caption{Four of the maps used in this analysis: the BLAST maps at 350 and 500\rmicron\ (\textit{top}) and the ACT data at 1380 and 2030 \rmicron\ for the same region (\textit{bottom}). Long wavelength modes in the ACT maps have been removed  using the high-pass filter described in eqn. (\ref{eqn:cos2filter}). All maps are multiplied by a taper as discussed in the text.  }
\label{fig:actblast}
\end{figure*}
% ======================

We compare to the models of \citet[hereafter B10]{bethermin2010} and \citet[hereafter M10]{marsden2010} for DSFGs at the BLAST and ACT wavebands.  These are phenomenological models which are specifically tailored to constrain the evolution of the rest-frame far-infrared peak of galaxies at redshifts up to $z\sim 4.5$.   The models use similar Monte Carlo fitting methods but differ in a few ways, e.g. the B10 model is fit using data (predominantly number counts) from a very wide range of IR wavelengths (15\rmicron\ to 1.1mm) while M10 uses only data that constrains the evolution of the FIR peak and was not fit to any observations with $\lambda < 70$\rmicron. B10 also divides galaxies into two distinct populations based on luminosity and attempts to account for the strong lensing of high-redshift galaxies; M10 does neither of the above.

% ======================
\subsection{Radio Galaxies}
\label{sec:radio}
Synchrotron and, to a lesser extent free-free emission dominates the SEDs of radio galaxies at rest-frame $\lambda \gsim 1~\rm mm$. Thus radio galaxies become an increasingly important contribution to the CSB at wavelengths greater than $\sim 1.5~\rm mm$  ($\nu \lsim 200 $ GHz).  Their number counts are relatively shallow \citep[e.g.,][]{dezotti2010}, meaning that their contribution to the power spectrum is dominated by the brighter sources, resulting in primarily Poisson noise, with the clustered term being sub-dominant.

While radio sources are a significant source of power at 2030 and 1380\rmicron, they do not feature prominently in the cross-frequency correlation of ACT and BLAST maps. They only contribute to the uncertainties in the cross-power spectrum, so  we do not include them in our models for the cross spectra.

\subsection{SZ}
\label{sec:sz}
The Sunyaev-Zeldovich effect \citep{sunyaev1972} is the distortion of the microwave background due to interaction of CMB photons with free electrons in clusters, and consists of two main physical mechanisms, referred to as \lq\lq thermal \rq\rq\ and \lq\lq kinetic\rq\rq.  The thermal term (tSZ) is the inverse Compton scattering of CMB photons as they pass through the intra-cluster medium in galaxy clusters.  The result is that the intensity of the CMB spectrum longward of 1380\rmicron\ (218 GHz) decreases (i.e., a decrement), while that shortward of 1380\rmicron\ increases (i.e., an increment), and 1380\rmicron\ is the null for the non-relativistic case.
The kinetic term (kSZ) is the Doppler shift of scattered CMB photons by the bulk motion of galaxy clusters.  The strength of the signal is proportional to the product of the free electron density and line of sight velocity.
% ======================
\subsection{Cirrus}
\label{sec:cirrus}
On large angular scales, a significant source of fluctuation power is emission from Galactic cirrus.  Although the SEP field is among the least contaminated by cirrus in the sky ($I_{\rm mean}=1.16~\rm MJy~sr^{-1}$), contributions from Galactic cirrus must still be accounted for.

The power spectrum of Galactic cirrus has been shown in many studies to exhibit power-law behavior.  Its amplitude varies over the sky, but its slope is always between $-2.6$ and $-3$ \citep[e.g.,][]{gautier1992, boulanger1996, miv2007,bracco2010}.  In the FIR/submm bands, the SED of Galactic cirrus is well described by a modified blackbody (Equation~\ref{eqn:bb}), with $T=17.5$~K and $\beta \sim 1.9$, peaking at $\lambda \sim 150$\rmicron\ \citep[V09,][]{bracco2010}.  As a result, bands closest to the peak (in our case 250\rmicron) are most susceptible to contamination.

As one moves far from the SED peak, \citet{finkbeiner1999} show that the modified blackbody approximation breaks down, and a multi-component fit is a much better description of the data.  Therefore, for this analysis we measure the power spectrum at 100\rmicron\ for the SEP region, and adopt model 8 of \citet{finkbeiner1999} to estimate the amplitude of the power in our bands.
% ======================
\section{Instruments and Observations}
\label{sec:data}
Below we describe the ACT and BLAST data used for the cross-correlation analysis, as well as the data used for estimating the cirrus power spectrum.  
\subsection{ACT}
\label{sec:ACT}
ACT is a 6-meter off-axis Gregorian telescope \citep{fowler2007} situated at an elevation of 5190 meters on Cerro Toco in the Atacama 
desert in northern Chile.  ACT has three frequency bands centered at 
\arone{} (2.0\,mm), \artwo{} (1.4\,mm) and \arthree{}  (1.1\,mm) with angular resolutions of roughly 1\farcm 4, 
1\farcm 0 and 0\farcm 9, respectively. The high altitude site in the arid desert is excellent for mm observations due to  low 
precipitable water vapor and stability of the atmosphere. The tropical location of ACT  permits observations on both  
the northern and southern celestial hemispheres. Further details on the instrument  are presented in 
\citet{swetz2010}, \citet{fowler2010} and references therein\footnote[1]{ACT Collaboration papers are archived at 
\url{http://www.physics.princeton.edu/act/}}.
The ACT maps used in this paper are made from the 2008 observing season data (\arone{} and \artwo{}, or 2030 and 1380\rmicron, respectively) and are identical to the maps used in \citet{hajian2010} and  \citet{das2010}. %Cross-correlations with BLAST are computed from a $\si m 9~\rm deg^2$ subset of the ACT southern strip which overlaps with the region observed by BLAST. 
The beam full widths at half maxima (FWHM) are $1.4\arcmin$ and $1.0\arcmin$ at \arone{} and \artwo{}, respectively \citep{hinks2010}. 
The maps have mean 1$\sigma$ sensitivities which vary slightly across the maps, ranging from $2.4-3.5~\rm mJy~ beam^{-1}$ (median $\approx 2.7~\rm mJy~ beam^{-1}$), and $3.2 - 5.4~\rm mJy~ beam^{-1}$ (median $\approx 3.7~\rm mJy~ beam^{-1}$), 
at \arone{} and \artwo{}, respectively \citep{das2010}. 
The map projection used is cylindrical equal area (CEA) with square pixels, $0.5\arcmin$ on a side.  The ACT data-set is divided into \ndivtext equal subsets in time, such that the \ndivtext  independent maps generated from these subsets cover the same area and have similar depths. 
We call these \lq\lq sub-maps\rq\rq.  
As described in \citet{hajian2010}, the ACT maps are directly calibrated to WMAP.  This results in a 2\% fractional temperature uncertainty for the \arone{} maps. The calibration error for the \artwo{} maps is 7\%.

Because the ACT maps have poorly measured modes on the largest angular scales,  we filter them using a
high-pass filter $F_{\rm c}(\ell)$ in Fourier space. The high-pass filter is a smooth sine-squared function in Fourier space given by
\be
\label{eqn:cos2filter}
F_{\rm c}(\ell) = \sin^2{x(\ell)} \Theta(\ell-\ell_{\mathrm{min}}) \Theta(\ell_{\mathrm{max}}-\ell) + \Theta(\ell-\ell_{\mathrm{max}}),
\ee
where $x(\ell)=(\pi/2)(\ell-\ell_{\mathrm{min}})/(\ell_{\mathrm{max}}-\ell_{\mathrm{min}})$ and $\Theta$ is the Heaviside function. We choose $\ell_{\mathrm{min}} = 100$ and $\ell_{\mathrm{max}}=500$. Moreover, the large-scale CMB in the ACT maps acts as noise in cross-correlations with the BLAST maps, since the CMB is absent in the latter.   If not ﬁltered, the large angular scale and CMB noise terms contaminate the  real-space cross-frequency correlations described in Section \ref{sec:real_space_xcorr}. Therefore we use a filter with $\ell_{\mathrm{max}}=2200$ when dealing with real-space cross-frequency correlations. The analyzed power spectra are corrected for this filter as well as for the effects of the beam and pixel window functions.

\begin{figure*}[ht]
\centering
\includegraphics[width=0.9\textwidth]{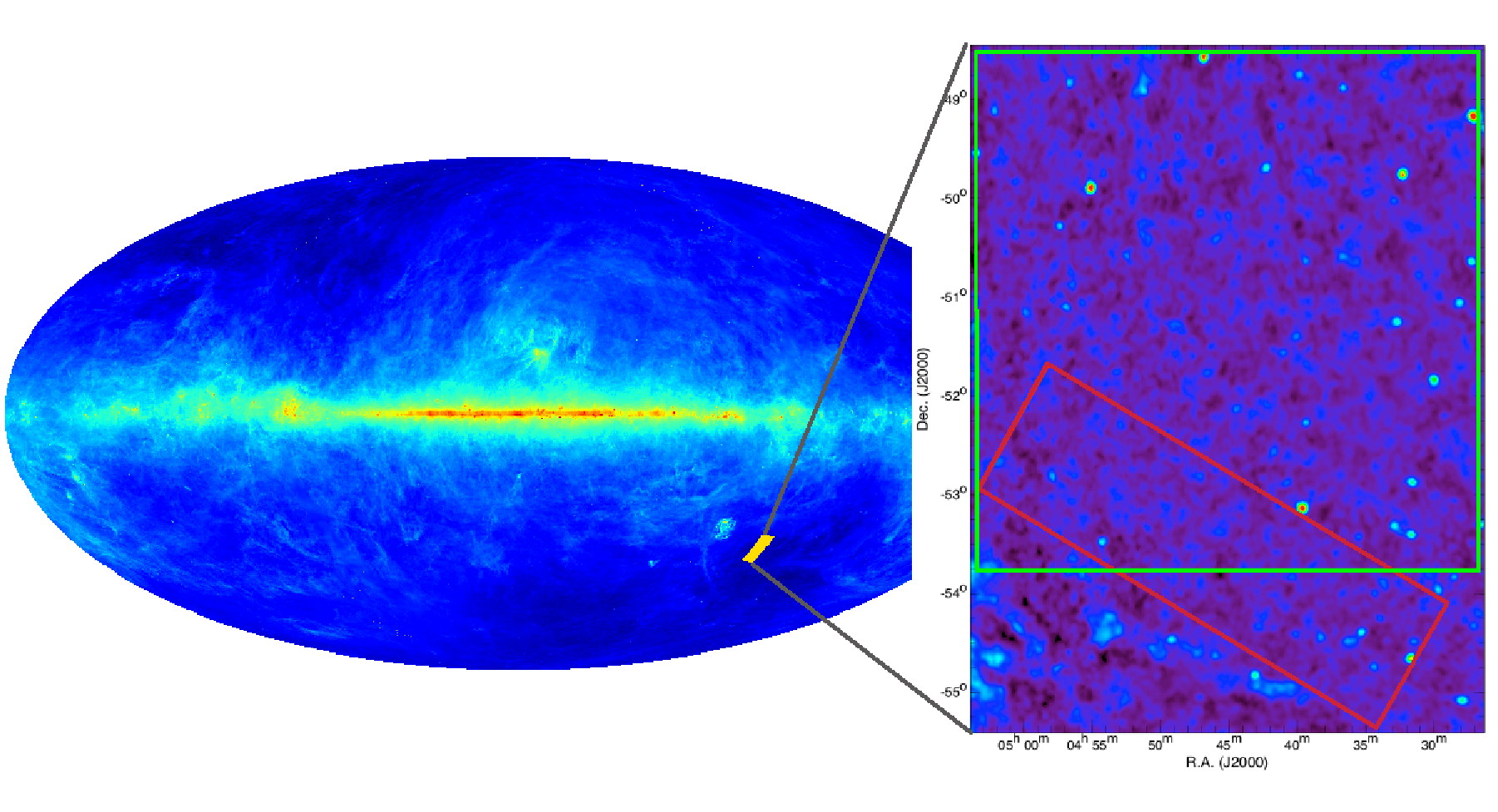}
\caption{The full-sky IRIS map at 100\rmicron\ scaled logarithmically.
The red outline represents the BLAST field in CEA projection which is used for cross-frequency correlating with the ACT maps.  The region clearly has low dust contamination. The bigger square area outlied by a green line shows the  $30~\rm deg^2$ region used for estimating the IRIS power spectrum. This map is filtered with the high-pass filter of Equation~\ref{eqn:cos2filter}. }
\label{fig:blastOnIRIS}
\end{figure*}

% ======================
\subsection{BLAST}
\label{sec:BLAST}
BLAST flew for 11 days from Antarctica in December of 2006.
Operating above most of the atmosphere, BLAST observed in bands which are difficult or not possible to observe from the ground.  As a pathfinder for the SPIRE instrument \citep{griffin2003}, it made observations at 250, 350 and 500\rmicron, of a number of targets, both Galactic and extragalactic.  Its 1.8-m under-illuminated primary resulted in beams with FWHM of 36, 45 and 60 arcseconds.  For a detailed description of the instrument see \citet{pascale2008} and \citet{truch2009}\footnote[2]{BLAST Results and Publications can be found at 
\url{http://blastexperiment.info/}}.

Among the fields BLAST observed is an 8.6 deg$^2$ rectangle near the South Ecliptic Pole (SEP) --- chosen because it is a relatively low-cirrus window through the Galaxy (see \S~\ref{sec:cirrus}) ---  whose corners lie at [ ($5^{\rm h} 02^{\rm m} , -52^{\circ} 50\arcmin$); ($4^{\rm h} 57^{\rm m} , -51^{\circ} 35\arcmin$); ($4^{\rm h} 25^{\rm m} , -54^{\circ} 19\arcmin$); ($4^{\rm h} 30^{\rm m} , -55^{\circ} 41\arcmin$)] (see Fig.~\ref{fig:actblast}).
Further studies of the SEP field, including BLAST catalogs and 24\rmicron\ maps, can be found in \citet{valiante2010} and \citet{scott2010}.
The maps have mean 1$\sigma$ sensitivities of 36.7, 27.2, and 19.1 mJy beam$^{-1}$ at 250, 350, and 500\rmicron, respectively.  Furthermore, confusion noise, due to multiple point sources occupying a single beam element, is estimated to be 7.6, 6.0 and 4.4 mJy beam$^{-1}$ at 250, 350, and 500\rmicron, respectively.  %
The 1$\sigma$ uncertainty on the absolute calibration is accurate to 9.5, 8.7, and 9.2\% at 250, 350, and 500\rmicron, respectively.

The BLAST time-ordered data (TODs) are divided into four sets --- covering the same region of the sky to the same depth --- from which we make four unique sub-maps.  The number of subsets is chosen to maximize the number of maps that can be made while maintaining uniformity in hits and providing as much cross-linking as possible.
The maps are made with the iterative mapmaker, {\tt SANEPIC} \citep{patanchon2009}, resulting in a transfer function of unity on the scales of interest.  
These sub-maps are unique to this study and are publicly available at {\tt \url{http://blastexperiment.info/results.php}}.  

Due to poor cross-linking, however, large scale noise, resembling waves in the map, is present.  This noise is easily dealt with by filtering in Fourier space, as described in \S~\ref{sec:formalism}.
Maps are made in tangent-plane projection (TAN), with 10\arcsec\ pixels.  In order to cross-correlate with ACT, BLAST maps are re-binned to ACT resolution and reprojected to cylindrical equal-area projection (CEA) using {\tt Montage}.\footnote[3]{\tt \url{http://montage.ipac.caltech.edu/}}  We confirm the alignment  by analyzing the real-space cross-frequency correlation, described in detail in \S~\ref{sec:real_space_xcorr}.

% ======================
\subsection{IRIS}
\label{sec:IRIS}
To estimate the contribution from Galactic cirrus, we use three IRIS \citep[reprocessed \emph{IRAS}:][]{miv2005} HCON\footnote[4]{HCON refers to each individual observation at three different epochs. For more information, and publicly available maps, see {\url{http://www.cita.utoronto.ca/~mamd/IRIS/}}} maps at 100\rmicron.  These maps are consistent with the  \citet[FDS]{finkbeiner1999} maps used for estimating the Galactic cirrus in \cite{das2010}, but they are at a higher resolution.
Since we are most interested in large-scale modes, the power spectrum is measured for a $30~\rm deg^2$ field surrounding the SEP (see Fig.~\ref{fig:blastOnIRIS}).  The three HCONs are from the same region of the sky, with independent noise properties. The power spectrum of the Galactic cirrus is computed from the average cross-spectrum of these three maps.  As a last step we correct the power spectra for a window function corresponding to IRAS's $4.3\arcmin$ beam.

% ======================

\subsection{Comparing Data sets: Testing Alignment with Real-Space Cross-Correlations}
\label{sec:real_space_xcorr}
\begin{figure}[b]
\centering
\includegraphics[width=1.05\linewidth]
{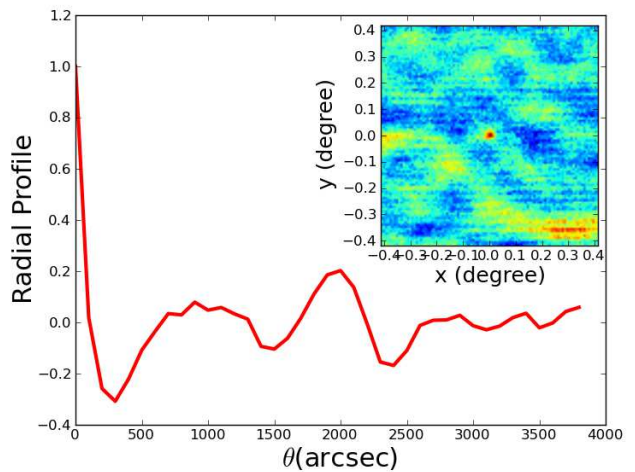}
\caption{The radial profile of the 2D cross-frequency correlation function between the ACT 1380\rmicron\  band and the BLAST  500\rmicron\ is plotted  with arbitrary normalization, with an image of the 2D function inset.  The function has a clear peak at zero lage.  
  This shows the two data sets are aligned and there is a correlated signal in the maps. Large scale fluctuations (at $\theta \ga 500 \arcsec$) are caused by the atmospheric noise and the CMB. The maps are filtered by the high-pass filter of Equation~\ref{eqn:cos2filter} with $\ell_{\rm min}=100$ and $\ell_{\rm max}=2200$ to remove longer wavelength noise.}   
\label{fig:align}
\end{figure}

% ======================
We test that the maps are properly aligned by inspecting their real-space cross-frequency correlations.  This is done by inverse Fourier transforming the 2D cross-frequency correlation of the Fourier components of the maps:
\begin{equation}
M_{a\times b}({\bf x}) = \sum_{{\mathbf{\ell}}} a({\bf \ell})b^*({\bf \ell})F_{\rm c}^2({\bf \ell}) {\rm exp}(i{\bf \ell \cdot x}),
\end{equation}
where ${\bf \ell}$ is a vector in Fourier space, $a({\bf \ell})$ and $b({\bf \ell})$ are the ACT and BLAST maps in Fourier space, respectively, and  $F_{\rm c}({\bf \ell})$ denotes high-pass filtering of Equation~\ref{eqn:cos2filter} in Fourier space. 
 The resulting measured real-space cross-frequency correlation function $M$  encodes the celestial correlation function between the bands,  ($C(x)$)  as: 
\begin{equation}
M_{a\times b}({\bf x}) = C({\bf x}) \otimes B({\bf x})  + n({\bf x}),
\end{equation}
where $\otimes$ denotes a convolution in real space, $B({\bf x})$ is the effective beam between the two maps, and $n({\bf x})$ is the noise.
Perfectly aligned maps would result in a 2D cross-frequency correlation function whose peak lies at ${\bf x} =0$.  The shape of $M_{a\times b}({\bf x})$  depends on the correlation length of the field, the high-pass filtering and the beams (See \citet{hajian2010} for a comparison  between measurements of a real-space cross-freqency correlation function and simulations for a related example).
As illustrated in Fig.~\ref{fig:align}, we measure a cross-correlation with a peak at ${\bf x} =0$ (zero lag),  indicating that the maps are correlated and properly aligned.

% ======================

\begin{figure*}[ht*]
\centering
\plottwo{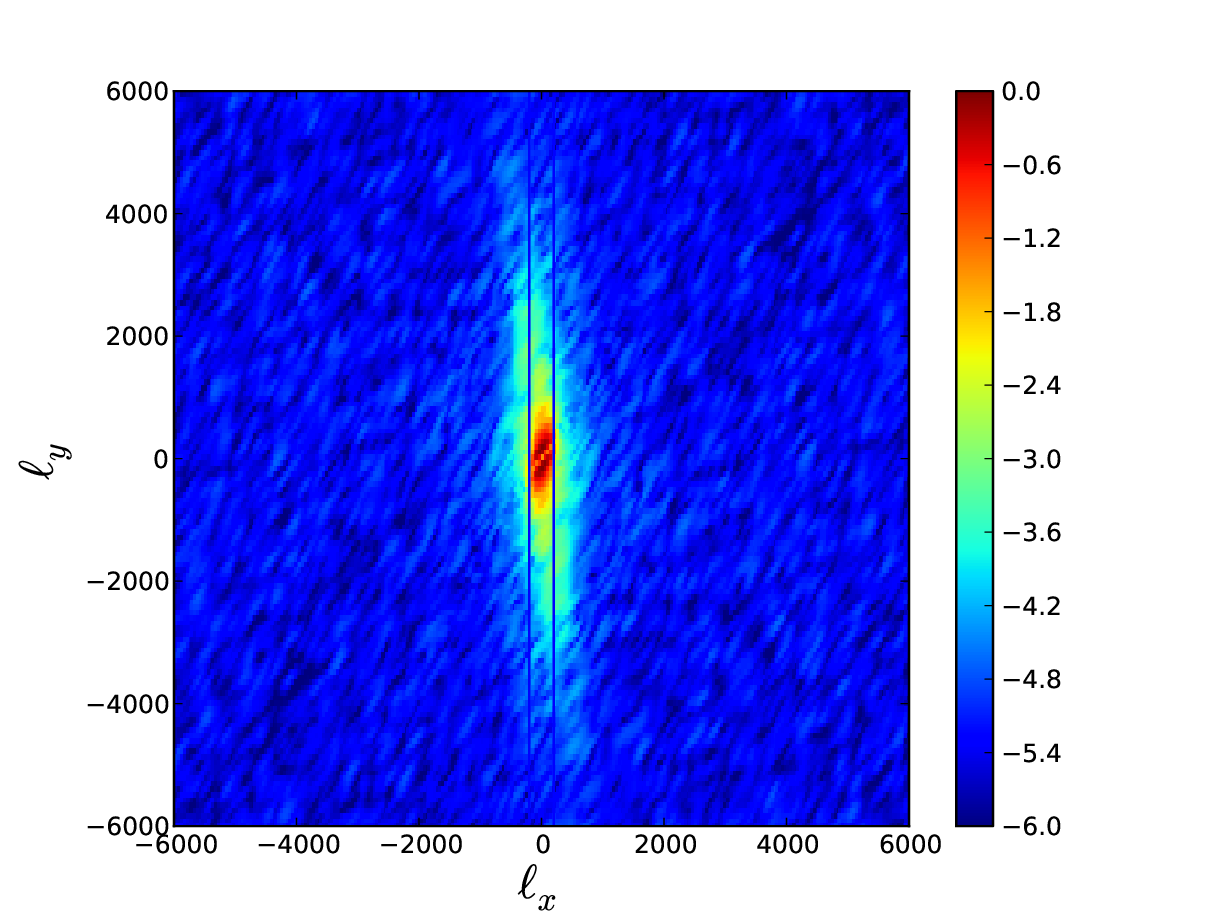}{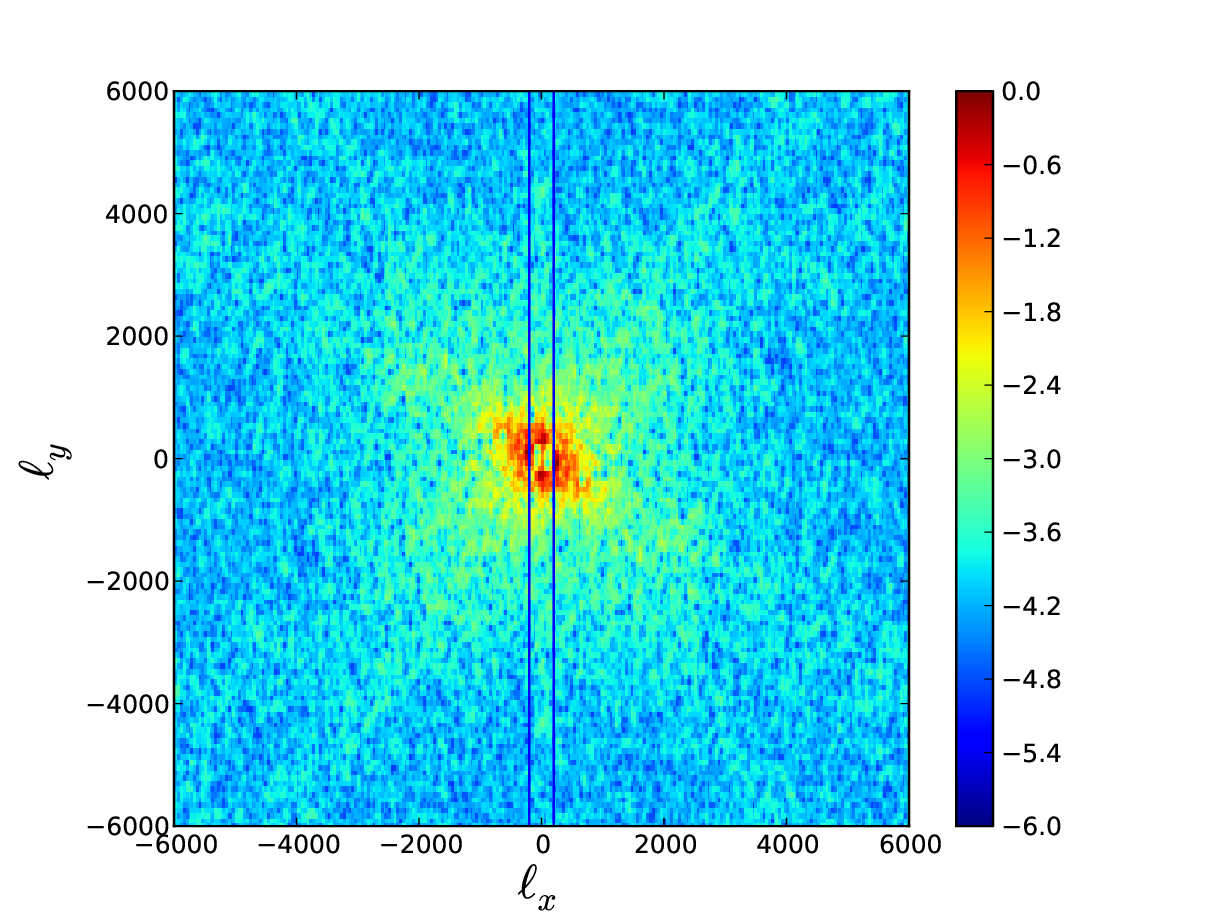}
%\subfigure[250\rmicron]{\label{fig:2dfft_BLAST}\vspace{0.1cm}\includegraphics[width=.47\linewidth]{250_2D.eps}}
%\subfigure[1380\rmicron]{\label{fig: 2dfft_act_blast}\vspace{0.1cm}\includegraphics[width=.47\linewidth]{220_2D.eps}}
\caption{The estimated 2D noise power spectrum for (a) $250~\mu $m and (b) $1380~\mu$m.   The noise model is computed from the difference of the auto- and cross-power spectra.  Noise due to scan-synchronous signals and other large-scale correlations, which contaminates the signal in the central region, is down-weighted with this weight map in Fourier space, and then further filtered with a vertical mask spanning $|\ell_x| < 500$. The weights are normalized by the maximum value of the noise power spectrum and are plotted in a log-scale. }
\label{fig:twoffts}
\end{figure*}

%\newpage
\section{Measuring Auto- and Cross-Frequency Power Spectra}
\label{sec:formalism}
We examine three distinct types of power spectra: the BLAST auto-band spectra; the BLAST cross-band spectra; and the ACT/BLAST cross-frequency spectra. We also estimate the auto-band spectrum for the 100\rmicron\ IRIS maps. Details of our power spectrum method are given below. 
\subsection{Power Spectrum Method}
We use three distinct types of power spectra: the BLAST auto-band spectra; 
the BLAST cross-band spectra; 
and the ACT/BLAST cross-band spectra. 

The maps can be represented as
\begin{equation}
\Delta T(\vec x) =  \Delta T_{\rm sky}(\vec x)\otimes B(\vec x) + N(\vec x),
\end{equation}
where $ \Delta T_{\rm sky}(\vec x)$ is the sky temperature signal, $N$  is the noise,  $B$ is the instrument beam and  we use $\otimes$ to represent a convolution in real-space.
Both the ACT and BLAST maps are made with unbiased iterative map-makers, whose transfer functions are approximately unity on the angular scales of interest in this study, and can thus be safely neglected.
% ======================

%%%%%%%%%%%%%%%%%%%%%%%%%%%%

All power spectra, both auto- and cross-frequency spectra, are computed using cross-correlations of sub-maps (described in \S~\ref{sec:ACT} and  \S~\ref{sec:BLAST}),
the advantage being that the noise between sub-maps is uncorrelated and thus averages to zero in the cross-spectra.  A cross-power spectrum computed this way provides an unbiased estimator of the true underlying power spectrum.
The power spectrum methods used in this paper closely follow those used for cross-correlating ACT and WMAP in  \cite{hajian2010}.

$\rm BLAST \times BLAST$ power spectra are computed using the average of the six  cross-correlations between the four BLAST sub-maps (in each band), such that
\begin{equation}
C_{\ell} = \frac{1}{6} \sum^{1 \le \beta \le 4}_{\alpha, \beta; \alpha < \beta} C_{\ell}^{\alpha \beta},
\end{equation}
where $\alpha$ and $\beta$ index the four sub-maps.  The reason for six cross-correlations, rather than nine, is that cross-frequency cross-correlations of sub-maps made from the same scans are not used, in order to avoid introducing correlated noise or other systematic effects.

Since the noise in all ACT and BLAST sub-maps is uncorrelated, we co-add the sub-maps for each frequency before computing the ACT$\times$BLAST power spectra.
The $\rm ACT\times BLAST$ power spectra are computed from these maps. This is identical to cross-correlating all ACT sub-maps with all BLAST sub-maps and averaging them.

Several components contribute to the cross-spectrum uncertainties. An analytic approach to computing the uncertainties is described in Appendix \ref{app:errorbars}.

%%%%%%%%%%% BEGIN FIGURE 4%%%%%%%%%%%%%%
\begin{figure*}[rt]
\centering
\includegraphics[width=1.\linewidth]{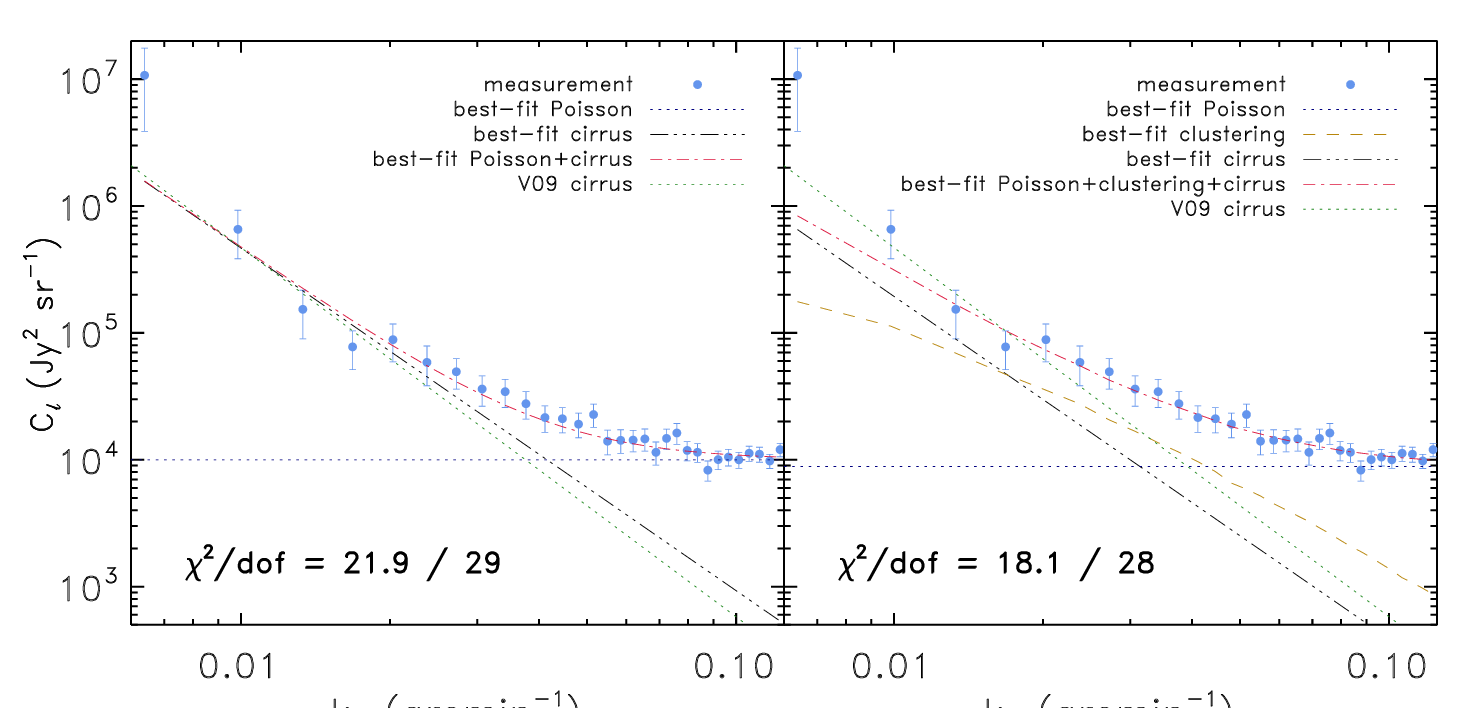}
\vspace{0.1cm}\caption{Power spectrum of the 30 deg$^2$ region surrounding the SEP from IRIS 100\rmicron\ maps.  A three component fit (left panel), including a clustering term (dashed yellow line) provides a better description of the measured power spectrum than fitting just Poisson and cirrus alone (right panel), but the difference is not statistically significant.}
\label{fig:cirrus_ps}
\end{figure*}
%%%%%%%%%% END FIGURE 4%%%%%%%%%%%%%%

\subsection{ Weights and Masks}
We use a flat-sky approximation for computing the power spectra, which are measured from Fourier transforms of the maps. Since the maps are 2D, the power spectra that we obtain are 2D.  
We adopt several techniques developed  in \cite{hajian2010} 
 in order to isolate and remove or down-weight instrumental and systematic noise. This is done in two stages, in real space and in Fourier space, before reducing the 2D spectrum into the familiar  1D angular power spectrum binned in radius ($\ell$).
 The $\rm ACT \times BLAST$ cross-spectra are limited by the area of the BLAST maps, which are $1.5 \times 5.7^{\circ}$ rectangles ($\sim 8.6$ deg$^2$), rotated by approximately $30^{\circ}$, and with noisy edges (see Fig.~\ref{fig:actblast}).  To apodize the sharp edges in the maps, we use the first Slepian taper \citep[properties of which are described in][]{das2009} defined on the BLAST region.  The gradual fall-off of this taper at the edges reduces the mode-mode coupling in the measured power spectrum.  
Any residual mode coupling caused by this weighting is corrected
in the end by deconvolving the window function (i.e. the power spectrum of the taper function) from the measured power spectrum using the algorithm described in \citet{hivon2002} and \citet{das2009}.  
Large-scale noisy modes are best treated in Fourier space.  The statistical isotropy of the Universe leads to an isotropic 2D power spectrum from extragalactic sources, on average.  
Anisotropic power in 2D Fourier space is caused by noise and is optimally dealt with using inverse noise weighting.  This is done by dividing the 2D spectra by our best estimate of the 2D noise power spectrum for each map. At each frequency band, the noise model is computed from the difference between the average 2D auto- and cross-spectra for each pair of maps as described in \cite{hajian2010}. For every cross-correlation, noise weights are computed from the inverse of the square root of the product of the two noise power spectra corresponding to the two frequency bands. The weights are whitened by dividing by their angular averaged value with a fine binning. Using simulations we confirm that this weighting does not bias the signal power spectrum.
Anisotropic, large-scale noise is evident at the center of the 2D power spectrum, as shown in Fig.~\ref{fig:twoffts}.  Noise-weighting down-weights the noisy vertical stripe that passes through the origin, which is predominantly due to large-scale unconstrained modes in the map \citep[see also][]{fowler2010}.  To further ensure that our results are not contaminated by this stripe, we mask a vertical band spanning $|\ell_x| < 500$ before  averaging the 2D power spectra in annuli.  Our results are invariant under further widening of this mask.
In order to be consistent with the V09 analysis, we make a mask to remove all point sources that have a flux greater than 0.5 Jy in the BLAST 250\rmicron\ map (six sources). 
We use that mask for computing BLAST power spectra only.   
For cross-correlations of ACT and BLAST, we instead mask just the brightest source in BLAST maps, which happens to be a local spiral galaxy.  Masking more point sources does not have an effect on the cross-spectra.  We also mask the known radio galaxies \citep{marriage/etal:prepa} in the ACT 2030\rmicron\ map to reduce the uncertainties in the cross-spectra (see Appendix \ref{app:errorbars}).

\subsection{Estimating Galactic Cirrus Emission}
\label{sec:cirrus_measurement}
IRIS maps at 100\rmicron\ contain three potential sources of power --  diffuse Galactic emission (or cirrus), as well as the Poisson and clustered terms of the DSFG power spectra.  Cirrus dominates the power spectrum on angular scales $\ell \gsim 800$, but varies depending on the observed patch of sky.  The  Poisson level is highly  sensitive to the adopted flux cut.  Thus, in order to detect the signal from clustering, both the cirrus and Poisson noise must be sufficiently low.  This is achieved by observing in a clean patch of sky, and cutting bright point sources.  We realize these criteria by considering only the SEP ($I_{\rm mean}=1.16~\rm MJy~sr^{-1}$), and by cutting sources with $S_{\rm cut} > 1~\rm Jy$.

\citet{miv2007} show that the power spectrum of Galactic cirrus can be approximated by a power-law,
\begin{equation}
P_{\mathrm{cirrus}}(k_{\theta}) = P_0\left( \frac{k_{\theta}}{k_0} \right) ^{\alpha},
\end{equation}
where $k_{\theta}$ is the angular wavenumber in inverse arcminutes, and $P_0$ is the power spectrum value at $k_0 \equiv 0.01$ arcmin$^{-1}$.  

%%%%%%%%%%% BEGIN FIGURE 5%%%%%%%%%%%%%%
\begin{figure*}[th]
\centering
\hspace{5.0cm}\vspace{0.1cm}\includegraphics[height=11.0cm,width=17.5cm]%[width=.90\linewidth]
{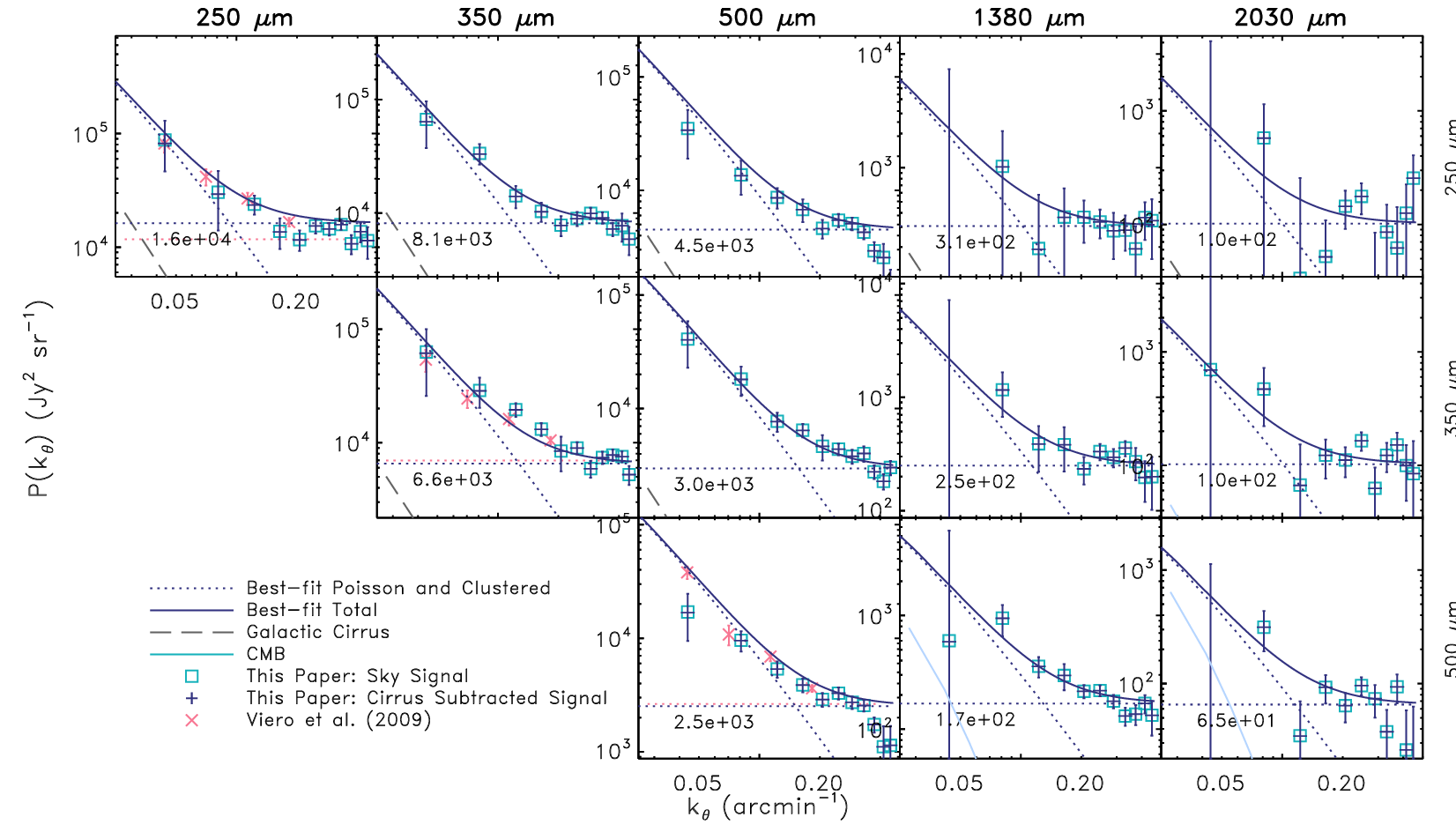}
\caption{$\rm BLAST\times BLAST$ (250--500\rmicron) and $\rm ACT \times BLAST$ (1380--2030\rmicron) power spectra in $P(k_{\theta})$ with 1$\sigma$ errorbars.   Squares and crosses are the data before and after cirrus removal.  Red exes and red horizontal dotted lines are published power spectra and Poisson noise levels from V09.  The dotted blue lines which are horizontal and which are falling with $k_{\theta}$, are the best-fit Poisson and clustering terms, respectively.  
  The departure from Poisson is the evidence for clustering of DSFGs.  
Note that the vertical scale is different for each panel.  The error bars are described in Appendix \ref{app:errorbars}.}
\label{fig:cl_spectra}
\end{figure*}
%%%%%%%%%% END FIGURE 5%%%%%%%%%%%%%%
We are only concerned with the modes that affect our measurement, and since the cirrus power spectrum falls steeply with increasing $k_{\theta}$, we focus our attention on the larger-scale modes.  
In order to probe these modes with maximum resolution in Fourier space, we measure the cirrus component of a $\sim 30~\rm deg^2$ region of the IRIS data surrounding the SEP field as indicated in Fig. \ref{fig:blastOnIRIS}.
We filter the IRIS maps using the high-pass filter of Equation~\ref{eqn:cos2filter}.  
The power spectrum is computed from  the mean of the three cross-spectra from the three HCON maps \citep*[using one taper at resolution 1; see][for details]{das2009}.

We attempt to fit the data in two ways, where in both cases we fix the slope of the cirrus power spectrum at $\alpha=-2.7$ \citep[adopting the properties of region 5 of][whose mean flux density most resembles the SEP]{bracco2010}.  The first is a two-parameter fit, where the free variables are the Poisson level and  the amplitude of the cirrus power.  For this we find $P_0 = (0.47 \pm 0.06) \times 10^6~\rm Jy^2~sr^{-1}$  and $\chi^2 =21.9~(\rm dof=29)$.  The measured power spectrum is shown in Fig. \ref{fig:cirrus_ps}. The power spectrum uncertainties are calculated in a manner analogous to that described in Appendix \ref{app:errorbars}.

The second fitting procedure uses a three-parameter fit in which the free variables are the Poisson level, cirrus amplitude and clustering amplitude of the DSFGs.  The shape of the clustering component is simply that of a linear dark matter spectrum.  In this case we find cirrus values $P_0 = (0.19 \pm 0.15) \times 10^6~\rm Jy^2~sr^{-1}$ and a clustering amplitude of  $\sim 720 \, \rm Jy^2~sr^{-1}$ at $\ell=3000$, with $\chi^2= 18.1~(\rm dof=28).$

The two approaches estimate consistent Poisson levels, but the fit with a clustered component appears to describe the data better than without, with $\Delta \chi^2 = 3.8$.  While not yet statistically significant, future studies with PACS at 100\rmicron\ \citep{poglitsch2010} should be able to measure the clustered signal to high significance.  
The cirrus power spectrum is assumed to continue to smaller angular scales, and is estimated at the ACT and BLAST bands using the average dust emission color $(I_{\mathrm{\rm (sub)mm}}/I_{100})^2$, which is estimated using model 8 of \citet{finkbeiner1999}.

\section{Power Spectrum Results}
\label{sec:results}
The BLAST auto-band and cross-band power spectra and $\rm BLAST\times ACT$ cross-frequency power spectra are shown in Fig.~\ref{fig:cl_spectra}.
Raw data are shown as squares, while cirrus subtracted points are shown as crosses with error bars. 
The Galactic cirrus spectra, interpolated to our bands as described in \S~\ref{sec:cirrus}, are shown as dashed lines in the bottom left corner of each panel (when strong enough to appear at all).  Cirrus appears to have a nearly negligible effect on the power in most bands, with only a marginal contribution in the 250\rmicron\ auto-spectrum.  Note, the cirrus contribution in V09 to the BLAST bands was extrapolated from 100\rmicron\ incorrectly; however, properly accounting for cirrus ultimately has little impact on the final result.  
The cirrus-corrected data are given in Table~\ref{tab:dl_data}. We describe the   models and the fits to these data in \S~\ref{sec:bias}.

The figure shows a clear cross-correlation between ACT and BLAST.  There is both a significantly correlated Poisson term (horizontal line) and a clear clustering term (rising to low $k_{\theta}$). 
This is the main result of this paper: that the unresolved BLAST background made up of DSFGs is intimately related to the ACT unresolved background. 
The signal is clearest in the ACT 1380\rmicron\  correlation with BLAST 500 and 350\rmicron, and less significant in the ACT 1380 and 2030\rmicron\ correlation with BLAST 250\rmicron.  
Additionally, the figure confirms the V09 BLAST power spectrum analysis, and extends it to include the cross-frequency correlation between BLAST bands.

Not shown in Fig.~\ref{fig:cl_spectra} are predictions for the cross-correlation of the SZ increment and decrement, nor that of predictions for the cross-correlations of the SZ decrement and DSFGs.  Both of these signals would appear as anti-correlated at the ACT 2030\rmicron\  band, and would act to decrease the total sky signal.  The former, using templates of \citet{battaglia2010}, was predicted to be negligibly small; and while at some level the latter should exist, we have not yet identified a clear signature (which should appear only in the cross-correlations with the ACT 2030\rmicron\  band).
%%%%%%%%%% BEGIN FIGURE 7%%%%%%%%
\begin{figure}[th*]
\centering
\includegraphics[width=.75\linewidth]{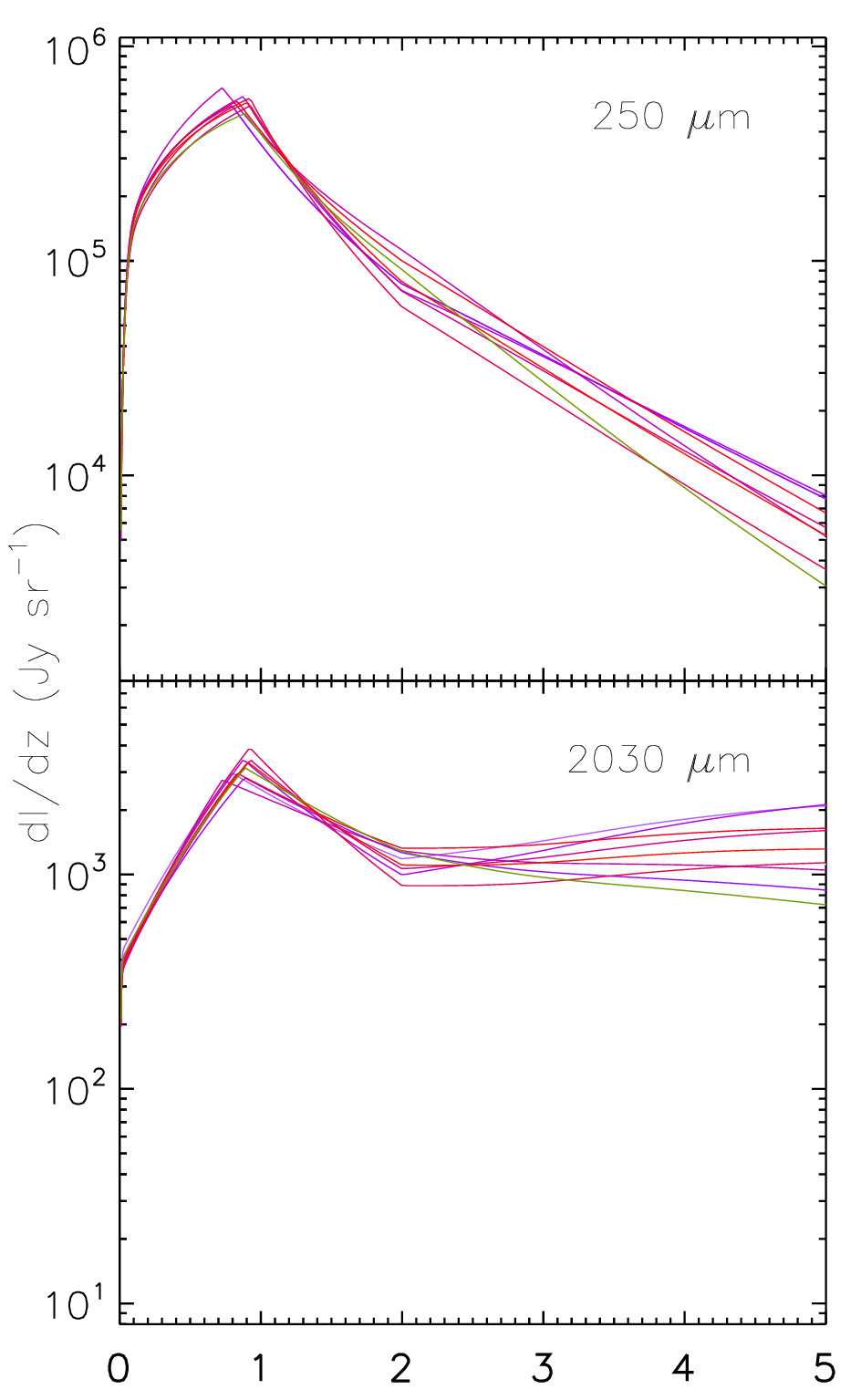}
\vspace{0.4cm}\caption{Redshift distributions of intensity, $dI/dz$, at 250 and 2030\rmicron\ for ten arbitrarily chosen realizations of the B10 source model. In this model the emission peaks around $z=1$ for the whole range of wavelengths covered by ACT and BLAST, but there are significant contributions to the IR flux from $z\gsim2$ at mm wavelengths.}
\label{fig:dsdz}
\end{figure}
%%%%%%%%%% END FIGURE 7 %%%%%%%%

%%%%%%%%%% BEGIN FIGURE 8%%%%%%%%
\begin{figure}[h*]
\centering
\includegraphics[width=1.0\linewidth]{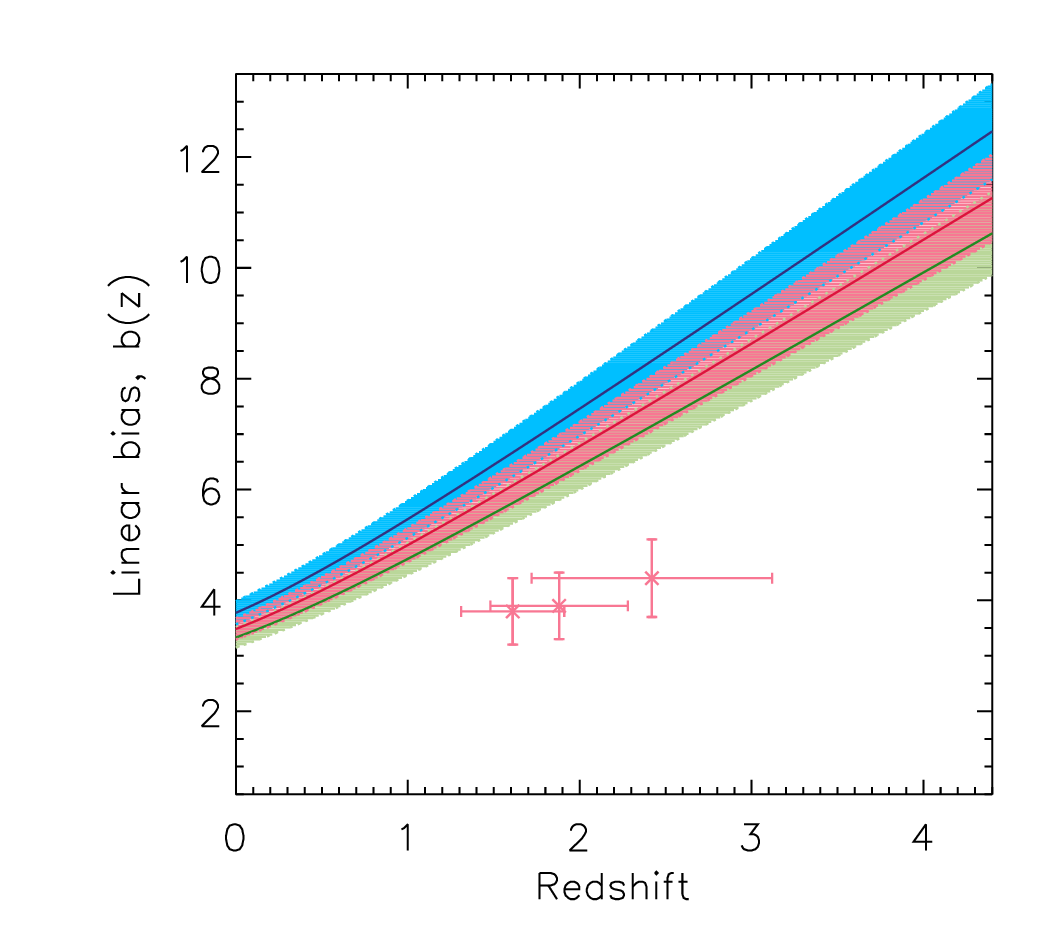}
\caption{Redshift-dependent best-fit linear bias $b(z)$ for three realizations of the B10 model with 1$\sigma$ error bounds estimated from our MCMC. Also shown are the best-fit single-value linear biases found in V09 for the BLAST 250, 350, and 500\rmicron\ auto-spectra, plotted at median redshifts $z= 1.61$, 1.88, and 2.42, respectively.  
}
\label{fig:bias}
\end{figure}
%%%%%%%%%% END FIGURE 8 %%%%%%%%

\section{Linear clustering model}
\label{sec:bias}
In this section we estimate the DSFG Poisson power levels, and fit the clustered component using a simple linear model similar to that of V09 and \citet{hall2010}.  We assume that the clustered component of the DSFG power spectrum, $P_{\rm DSFG}$, is related to the linear dark matter power spectrum, $P_{\rm DM}$, through a single bias parameter $b(z)$:
\begin{equation}
P_{\rm DSFG}(k,z)=b(z)^2P_{\rm DM}(k,z),
\end{equation}
so that the angular power spectrum, $P(k_{\theta})$, of DSFGs can be written as
\begin{eqnarray}
P_{\nu_1,\nu_2}(k_{\theta})=&\int dz \left(\frac{dV_{\rm c}}{dz}(z)\right)^{-1}b^2(z)P_{\rm DM}\left( \frac{2 \pi k_{\theta}}{x(z)},z\right)\times \nonumber \\
&\frac{dI_{\nu_1}}{dz}(z)\frac{dI_{\nu_2}}{dz}(z),
\end{eqnarray}
\citep{BCH2, tegmark2002}, where $x(z)$ is the comoving distance, $dV_{\rm c}/dz=x^2dx/dz$ the comoving volume element, and $dI/dz$ is the contribution to the intensity from sources at redshift $z$. 

We adopt $dI/dz$ from the source model of B10.  Fig.~\ref{fig:dsdz} shows plots of $dI/dz$ from the model at 250 and 2030\rmicron\ for ten randomly chosen realizations provided by the B10 distribution,\footnote[5]{\url{http://www.ias.u-psud.fr/irgalaxies/model.php}}.  
We adopt the concordance model, a flat $\Lambda$CDM cosmology with $\Omega_{\rm M}=0.274$, $\Omega_{\Lambda}=0.726$, $H_0 =70.5$ km s$^{-1}$ Mpc$^{-1}$, and $\sigma_8 = 0.81$ \citep{hinshaw2009}.

The linear dark matter power spectrum is calculated as $P_{\rm DM} (k) = P_0(k) D^2 (z)T^2(k)$, where $P_0(k_{\theta})$ is the primordial power spectrum, $T(k_{\theta})$ is the matter transfer function with fitting function given in \citet{eisenstein1998}, and $D(z)$ is the linear density growth function.
For simplicity we treat the magnitude of the Poisson component of each of the 12 power-spectra as a free parameter rather than simultaneously using the B10 model to predict the Poisson level as well as the clustering power.

We note that our model does not account for non-linear, one halo clustered power.  Though likely present, the data are not sufficiently constraining given that the Poisson power dominates over or is degenerate with the one-halo term in the angular scales to which we are sensitive.
%%%%%%%%%% BEGIN FIGURE 9 %%%%%%%%
\begin{figure*}[th*]
\centering
\hspace{5.0cm}\vspace{0.1cm}
\includegraphics[height=11.0cm,width=17.5cm]{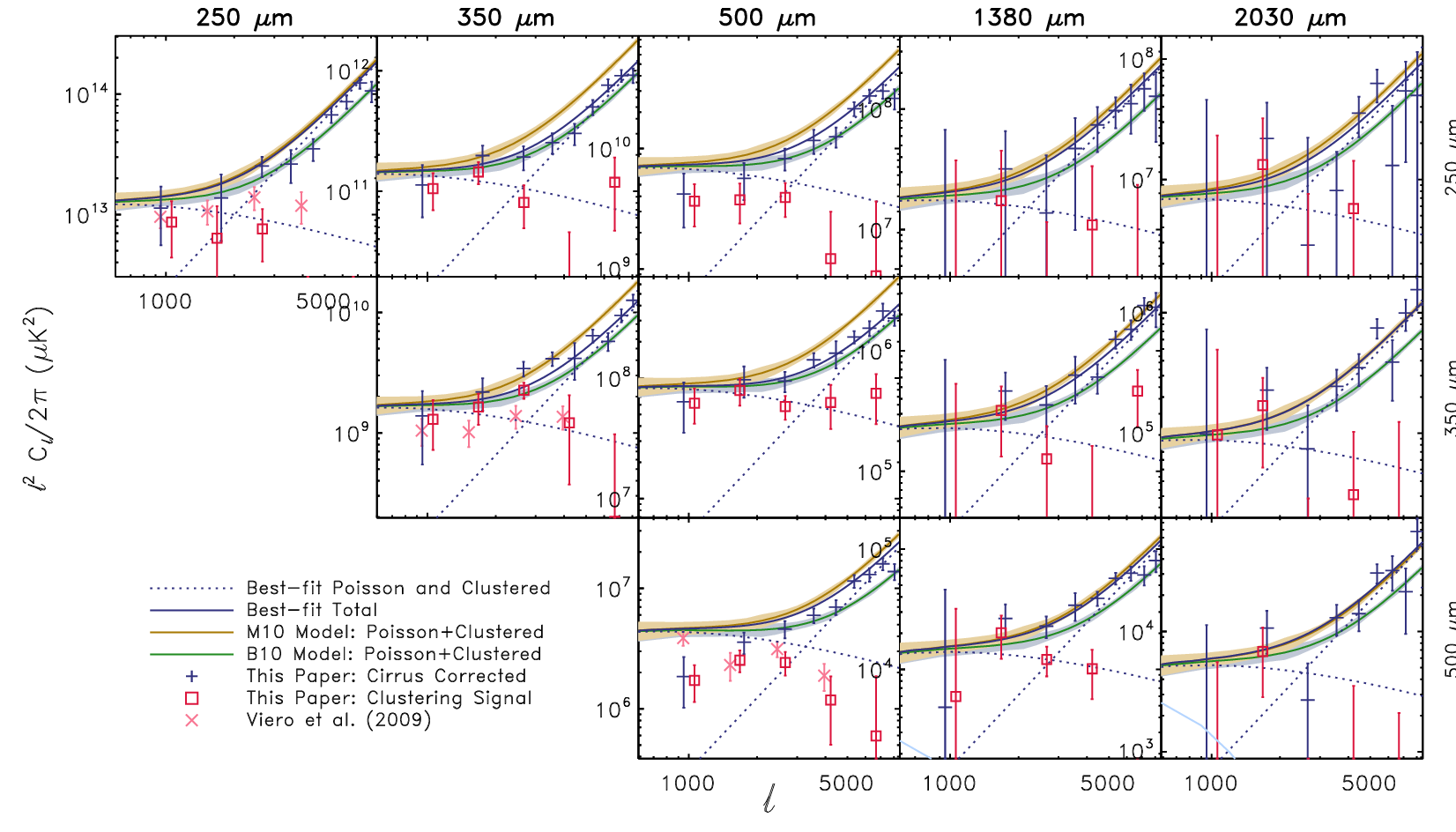}
\caption{$\rm BLAST\times BLAST$ (250--500\rmicron) and $\rm ACT \times BLAST$ (1380--2030\rmicron) power spectra in $\ell^2C_{\ell}/2\pi$.  Data, which have had Galactic cirrus power removed, are shown as blue crosses.   Red squares are the same data after removal of the Poisson term, and after logarithmic binning, with $\rm log(\Delta \ell) = 0.2$, and represent the contribution to the total power spectrum from clustering.  Pink exes are the clustered term data from V09.
The blue dotted lines rising to larger $\ell$ are the best-fit Poisson terms, and the approximately horizontal blue dotted lines are the best-fit clustering terms, which are determined by the $z$-dependent bias, as described in \S~\ref{sec:bias}.  
Also plotted are the phenomenological models of B10 and M10, in green and brown (with shaded error regions), respectively.  Poisson levels are calculated after truncating the counts at 300, 250, 170, 20, and 20 mJy at 250, 350, 500, 1380 and 2030\rmicron, respectively.  
 Error regions are calculated with Monte Carlos.  
 Both the models agree at some effective wavelengths, but disagree at others, so that neither describes the data fully.  
The M10 model also somewhat over-predicts the CIB at BLAST wavelengths, which is consistent with the behavior of the model Poisson term here.  %
Note that the vertical scale is different for each panel.  
The cirrus-corrected data here are given in Table~\ref{tab:dl_data}.
}
\label{fig:bl_spectra}
\end{figure*}
%%%%%%%%%% END FIGURE 9 %%%%%%%%
\subsection{Estimating the bias}
\label{EstimatingBias}
In principle the bias, $b$, is a function of scale and redshift, as well as environmental factors such as the host halo mass for the DSFGs. Here we adopt a simple redshift-dependent bias of the form \citep*{BCH2, hui2008}:
\beq
b(z) = 1+ (b_0 -1) \frac{D(z_0)}{D(z)} ,
\label{eqn:bz}
\eeq
where $D(z)$ is the linear growth function and $b_0$ is an initial bias at some formation redshift, $z_0$. This parameterization assumes that DSFGs are members of a single population, which formed at the same epoch ($z_0$) and under the same conditions.

Our parameter space consists of the 12 Poisson levels plus $b_0$ and $z_0$.  However, just as the Poisson contribution is a sum over the galaxy distribution, weighted by the square of the fluxes, eqn. \ref{eqn:shot_level}, so is the average bias, though weighted linearly by the flux \citep[e.g.,][]{bond93}. Thus, in any physical model for the star forming objects and their bias the two terms would be correlated, but in a model-dependent way, so here we just adopt the independent bias for simplicity. Decoupling the Poisson level and the clustered component means the interpretation of our derived $b$ is not straightforward. We find that moderate changes in $z_0$ (in the range $6<z_0<10$) have virtually no effect on the quality of the fit or best-fit $b(z)$, and also find that $b_0$ and $z_0$ are almost degenerate, and so we fix $z_0=8$.  We explore the remaining 13-parameter space using a Markov Chain Monte Carlo (MCMC) method with uniform priors on each parameter.  We fit the ACT$\times$BLAST data in the range $\ell = 950$ ($k_{\theta}=0.04~\rm arcmin^{-1}$).
We find a best-fit $b_0=18.2^{+2.3}_{-1.7}$ with $\chi^2 = 107$ for 101 degrees of freedom. 
This corresponds to $b(z=1)=5.0^{+0.6}_{-0.4}$ and $b(z=2)=6.8^{+0.8}_{-0.5}$.  
This Poisson plus clustering model is preferred to the null case with no ACT$\times$BLAST correlation at over 25$\sigma$. The best-fit clustering and Poisson levels are reported in Table~\ref{tab:c_l}, and are plotted as blue dotted lines in Figs.~\ref{fig:cl_spectra} and \ref{fig:bl_spectra}.

We also try fitting just the ACT$\times$BLAST data with no clustering power.  In this case we obtain a best-fit $\chi^2 =  64.3$ with 48 degrees of freedom. 
After adding linear clustering to the model, we find $\chi^2 = 43.6$ with 47 degrees of freedom,
corresponding to $\Delta \chi^2$ of 20.7 (with one fewer degree of freedom), so that the model including clustering is preferred to one with only Poisson and cirrus at greater than 4$\sigma$.  Additionally, the Poisson levels are lower when clustering is included.  

\begin{figure*}[th]
\centering
\vspace{0.2cm}\includegraphics[width=1.0\linewidth]
%{new_hall_plot.eps}
{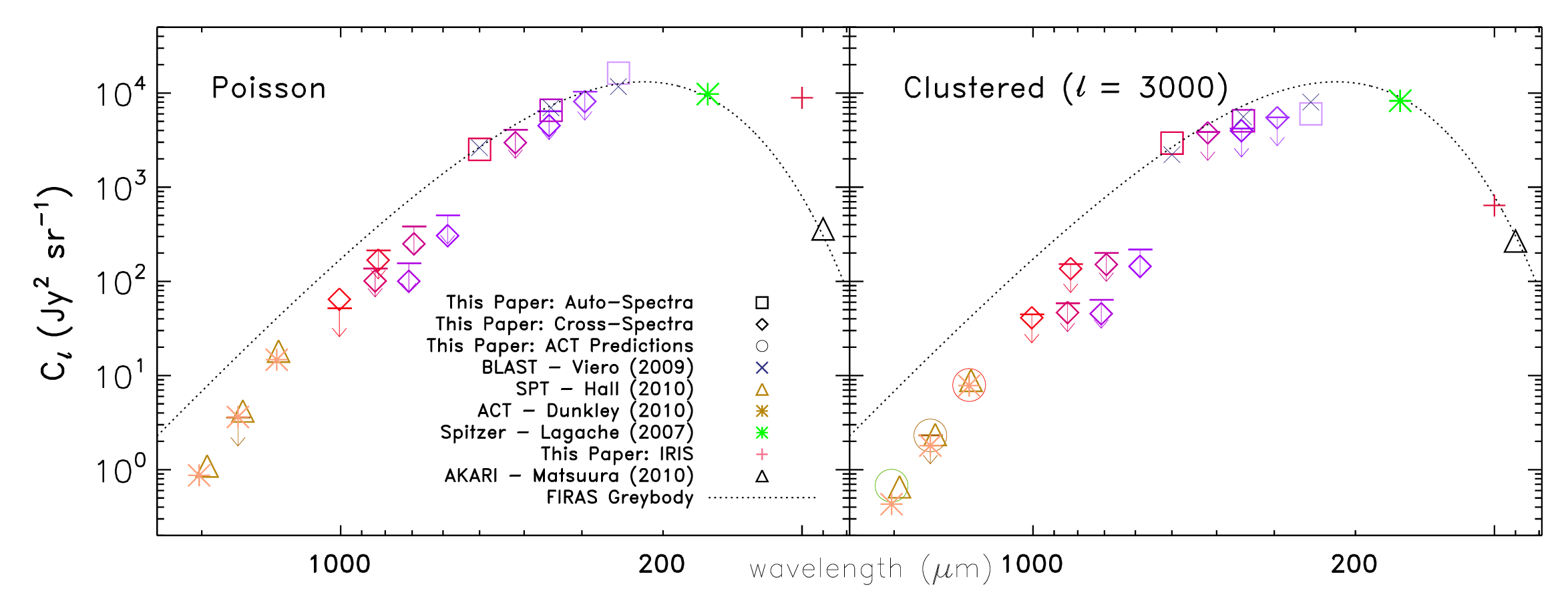}
\caption{$C_{\ell}$ versus wavelength for observations and models.  From left to right, the actual or effective wavelengths, $\lambda_{\rm eff} = \sqrt{\lambda_1 \lambda_2}$, (in\rmicron) are: 2030, 1673, 1380, 1007, 843, 831, 712, 695, 587, 500, 418, 354, 350, 296, 250, 160, 100, and 90.
Best-fit Poisson (left panel) and clustered (at $\ell = 3000$, right panel) $C_{\ell}$ from measurements are shown as squares (auto-spectra) and diamonds (cross-spectra), respectively, and our measurement of IRIS galaxies are shown as red crosses.  
Open circles represent the prediction for the clustered power at the ACT wavelengths from the best-fit, redshift-dependent bias model.  
Uncertainties are omitted for visual clarity, but are generally smaller than the size of the symbols due to the large dynamic range in $C_{\ell}$.     
The geometric mean of the cross-band spectra, defined as $\sqrt{C_{\ell}^{\lambda_1} \cdot C_{\ell}^{\lambda_2}}$,  are shown as downward-pointing  arrows.   %Note that, in order to make a fair comparison, the $\rm ACT\times BLAST$ Poisson geometric means are calculated from auto-frequency correlation levels derived with the $\rm ACT\times BLAST$ mask.  
Measurements from other experiments are: ACT \citep[yellow asterisks]{dunkley2010}; BLAST \citep[black exes]{viero2009}; \emph{Spitzer} \citep[green asterisk]{lagache2007}; SPT \citep[yellow triangles]{hall2010}; \emph{AKARI} \citep[black triangle]{matsuura2010}.  
The FIRAS modified blackbody ($T=18.5, \beta=0.64$) is plotted as a dotted line.  As was seen in \citet[Fig.~5]{hall2010}, FIRAS describes the data short of 500\rmicron, but over-predicts the measurements at millimeter wavelengths. The ratio of the measurement (diamonds) to the geometric mean (downward-pointing arrows) represents the level of cross-correlation between bands.  
}
\label{fig:nu_vs_cl}
\end{figure*}

Lastly, we try fitting a single-value bias, independent of redshift, to the entire range of power spectra.  We find a best-fit single-value bias of $5.0\pm0.4$ with $\chi^2=110.8$ for 101 degrees of freedom; worse than a redshift-dependent bias by $\Delta \chi^2=3.8$. The single-value bias thus provides a good fit to the ACT$\times$BLAST data, however when we include the measured 2030\rmicron\ (AR1) and 1380\rmicron\ (AR2) clustered power from \citet{dunkley2010} in the likelihood calculation, we find the redshift-dependent bias is preferred at $\sim 2 \sigma$. The ACT clustering measurements are reproduced well with $b(z)$ (see Fig. \ref{fig:nu_vs_cl}, right panel) but under-predicted with the single-value bias.

Fig.~\ref{fig:bias} shows $b(z)$ calculated using Equation~\ref{eqn:bz} for three realizations of the B10 model which span the entire range of results. The bias appears high compared to the linear bias estimates in V09 \citep[who adopted the][model]{lagache2004}, although given the spread in $b(z)$ from different realizations of the B10 model, the measurements are not inconsistent.  

Our choice of source model and bias parameterization is likely affecting $b(z)$. The B10 model contains two distinct classes of IR sources, \lq\lq normal\rq\rq\ and \lq\lq starburst,\rq\rq\ with substantially different luminosities, which is not accounted for in our $b(z)$ parameterization. Also, the B10 model does not match observational constraints equally well across the whole wavelength range probed by ACT and BLAST; for instance it under-predicts Poisson power and number counts compared to BLAST and SPT measurements at 500 and 1360\rmicron\ \citep[see \S~5.6 of][as well as Fig.~\ref{fig:bl_spectra} of this paper]{bethermin2010}. An under-prediction of $dI/dz$ would result in a higher bias to compensate.

%%%%%%%%%% BEGIN FIGURE 10 %%%%%%%%
%%%%%%%END FIGURE 10%%%%%%%%%%%%%
\section{Discussion}
\label{sec:discussion}
We compare the models of B10 and M10 (described in \S~\ref{sec:dsfg},) to our results.  
Poisson levels are calculated from the number counts using Equation~\ref{eqn:shot_level}, where the values of $S_{\rm cut}$ are chosen to mimic those used in the analysis, i.e., $S_{\rm cut} = 500$, 250, 170 mJy at 250, 350, and 500\rmicron, respectively. 

Model predictions of B10 and M10 are shown in Fig.~\ref{fig:bl_spectra} as green and brown lines with shaded error regions, respectively.  
Both models agree at some effective wavelengths, but disagree at others, so that neither appears to fully describe the data.  
The M10 model also somewhat over-predicts the CIB at BLAST wavelengths, which is consistent with the behavior we see for the model Poisson term.  %
On the other hand, as already mentioned, the B10 model under-predicts Poisson power and number counts compared to BLAST and SPT measurements at 500 and 1360\rmicron, which again is consistent with the behavior of the Poisson term of the model in Fig.~\ref{fig:bl_spectra}.

%%%%%%%%%% BEGIN FIGURE 11 %%%%%%%%%
\begin{figure*}
\centering
\vspace{0.2cm}\includegraphics[width=1.0\linewidth]{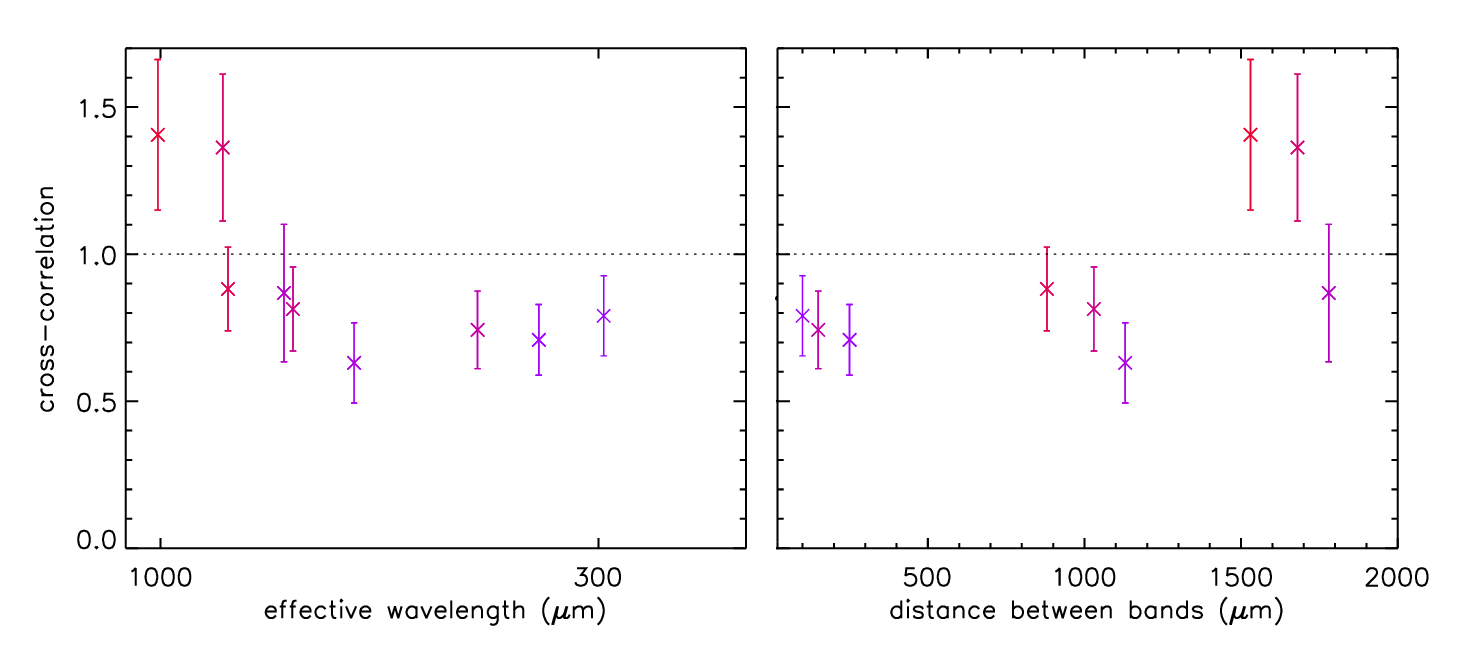}
\vspace{0.2cm}\caption{Left panel: Cross-frequency correlation versus effective wavelength for Poisson.  Cross-frequency correlation is defined as $C_{\ell}^{\lambda \lambda^{\prime}}/(C_{\ell}^{\lambda} \cdot C_{\ell}^{\lambda^{\prime}})^{1/2} $, i.e., the ratio of the measurement (diamonds in Fig.~\ref{fig:nu_vs_cl}) to the geometric mean (downward pointing arrows in Fig.~\ref{fig:nu_vs_cl}).  
%Note that, as in Fig.~\ref{fig:nu_vs_cl}, the $\rm ACT\times BLAST$ Poisson level geometric means are calculated from auto-frequency correlation levels derived with the $\rm ACT\times BLAST$ mask.  
From left to right, the actual or effective wavelengths, $\lambda_{\rm eff} = \sqrt{\lambda_1\times \lambda_2}$, (in \rmicron) are: 2030, 1673, 1380, 1007, 843, 831, 712, 695, 587, 418, 354, and 296.
The horizontal line at unity represents 100\% cross-correlation.  
Right panel: Cross-frequency correlation versus distance between bands.  Correlation is seen between all the ACT and BLAST frequencies.
}
\label{fig:dist_vs_cl}
\end{figure*}
%%%%%%%END FIGURE 11%%%%%%%%%%%%%

The Poisson and clustered power amplitudes are plotted as a function of effective wavelength,  defined as $\lambda_{\rm eff} = \sqrt{\lambda_1 \lambda_2}$, in Fig.~\ref{fig:nu_vs_cl}.  Also included are measurements made by the following experiments: \emph{AKARI} at 90\rmicron\ \citep[]{matsuura2010}; \emph{Spitzer} at 160\rmicron\ \citep[]{lagache2007};  BLAST at 250, 350, and 500\rmicron\ \citep[]{viero2009}; ACT at 1380, 1673, and 2030\rmicron\ \citep{dunkley2010}; SPT at 1363, 1629, and 1947\rmicron\ \citep[]{hall2010}; and the FIRAS modified blackbody ($T=18.5, \beta=0.64$), which is shown as a dotted line.

The degree of correlation between widely spaced wavelengths is of interest both in determining the redshift distribution of sources, and for modeling the IR source power as a CMB contaminant. To assess the correlation, the geometric means at each effective cross-band wavelength, defined as $\sqrt{C_{\ell}^{\lambda} \cdot C_{\ell}^{\lambda^{\prime}}}$, are shown as a downward pointing arrows. Since we do not measure $C_{\ell}^{\lambda}$ at $\lambda = 1380$ and 2030\rmicron, we rely on measurements by \citet{dunkley2010} for those bands when calculating the geometric means. 
The ratios of the measurements to the geometric means, $C_{\ell}^{\lambda \lambda^{\prime}}/(C_{\ell}^{\lambda} \cdot C_{\ell}^{\lambda^{\prime}})^{1/2} $, then represent the levels of cross-correlation between bands. These are shown in Fig.~\ref{fig:dist_vs_cl} for the Poisson power as a function both of effective wavelength and of distance between bands. Correlation is seen between all the frequencies, and does not fall significantly as a function of increased band separation, suggesting a tight redshift distribution for the overlapping population. This behavior is consistent with the findings of e.g., \citet{hall2010} and \citet{dunkley2010}: that the $1000-2000$\rmicron\ bands are correlated at close to the 100\% level, and extends the range of wavelengths probed.

\section{Conclusion}
\label{sec:conclusion}
We present measurements of the auto- and cross-frequency correlations of BLAST (250, 350 and 500\rmicron) and ACT (1380 and 2030\rmicron) maps.  We find significant levels of correlation between the two sets of maps, indicating that the same DSFGs that make up the unresolved fluctuations in BLAST maps are also present in ACT maps.  Furthermore, we confirm previous BLAST analyses \citep{viero2009} for a different field and with an independent pipeline, and extend the analysis by including $\rm BLAST\times BLAST$ cross-frequency correlations.  %

We fit Poisson and clustered terms at each effective wavelength simultaneously, which we achieve by adopting a model for the sources \citep{bethermin2010}, assuming a parameterized form for the $z$-dependent bias and using an MCMC to minimize the $\chi^2$.  Using this model we detect a clustered signal at $4\sigma$, in addition to a Poisson component.  The best-fit bias is one that increases sharply with redshift, and is consistent with what was found by \citet{viero2009}.  

We compare phenomenological models by \citet{bethermin2010} and \citet{marsden2010} to the data and find rough agreement at numerous effective wavelengths.  But we also find that neither model quite reproduces the data faithfully.  Thus, we expect this measurement and others like it will ultimately provide powerful constraints for the redshift distribution and SEDs of future versions of the models.

Though we find convincing evidence for correlated Poisson and clustered power from DSFGs, the levels of precision needed to robustly remove these signals from CMB power spectra demand better measurements still.  This is particularly true of the clustering term, whose contribution to the power spectrum in $\ell^2C_{\ell}$ peaks at $\ell \sim 800$ -- 1000, which is also the region in $\ell$-space typically targeted in searches for the SZ power spectrum.  Since the clustered term should scale independently of the Poisson term, the measurement becomes increasingly important to determine precisely.  
Future studies combining \emph{Herschel}/SPIRE with ACT, SPT, and \emph{Planck} will go a long way towards solidifying this much needed measurement.

% ======================
%%%%%%%%%% BEGIN TABLE 1 %%%%%%%%
\begin{table*}
  \centering
  \begin{tabular}{c|c||c|c|c|c|c}
   \hline
 $ $ & band ($\mu \rm m$) & 250 & 350 & 500 & 1380 & 2030 \\
   \hline
   \hline
 $ $ & $ 250$ &  $(1.1\pm0.1)\times10^{7}$ & $ (9.1\pm 0.6)\times10^{4}  $ &  $(3.1\pm 0.3)\times10^{3} $&  $ (1.\pm0.4)\times10^{1}$ & $(5.9\pm1.9)\times10^{0}$\\

 $ C_{\ell}^{p}~(\mu K^2) $ & 350 &  -- & $(1.1\pm0.1)\times10^{3}$ & $ (3.4\pm 0.3)\times10^{1}  $ &  $(1.8\pm 0.3)\times10^{-1} $&  $ (1.0\pm0.2)\times10^{-1}$ \\

  $ $ & $ 500  $ &  -- & -- & $ (1.8\pm 0.1)\times10^{0}  $ &  $(7.3\pm 1.0)\times10^{-3} $ &  $ (4.6\pm0.7)\times10^{-3}$ \\
   \hline
   \hline
% $ $ & $ 250$ &  $(1.12\pm0.31)\times10^{7}$ & $ (1.10\pm 0.31)\times10^{5}  $ &  $(4.40\pm 1.25)\times10^{3} $&  $ (2.16\pm0.60)\times10^{1}$ & $(7.43\pm2.10)\times10^{0}$\\
 $ $ & $ 250$ &  $(6.1\pm 1.1)\times10^{6}$ & $ (7.3\pm 1.1)\times10^{4}  $ &  $(3.6\pm 0.3)\times10^{3} $&  $ (9.3\pm1.1)\times10^{0}$ & $(3.6\pm0.4)\times10^{0}$\\

$ $ & 350 &  -- & $(8.7\pm 1.2)\times10^{2}$ & $ (4.4\pm 0.6)\times10^{1}  $ &  $(1.2\pm 0.2)\times10^{-1} $&  $ (4.7\pm0.7)\times10^{-2}$ \\

  $ C_{\ell=3000}^{c}~(\mu K^2) $ & $ 500  $ &  -- & -- & $ (2.1\pm 0.3)\times10^{0}  $ &  $(6.9\pm 1.1)\times10^{-3} $ &  $ (2.6\pm0.4)\times10^{-3}$ \\
  
  $ $ & $ 1380  $ &  -- & -- & -- &  $(3.4\pm 1.0)\times10^{-5} $ &  $ (1.2\pm0.3)\times10^{-5}$ \\
  
  $ $ & $ 2030  $ &  -- & -- & --&  -- &  $ (4.4\pm1.2)\times10^{-6}$ \\

\hline
\end{tabular}
\caption{Best-fit $C_{\ell}^{\rm Poisson}$ and $C_{\ell}^{\rm clustering}(\ell=3000)$, including predictions for the clustered power at the three effective ACT bands.  Predictions for the Poisson power at ACT bands are not provided as the Poisson terms are treated as free parameters when obtaining the best fit (see \S~\ref{sec:bias}).} 
  \label{tab:c_l}
\end{table*}
%%%%%%%%%% END TABLE 1 %%%%%%%%
%%%%%%%%%% BEGIN TABLE 2 %%%%%%%%
\begin{sidewaystable*}
  \begin{tabular}{c||c|c|c|c|c|c|c|c|c}
   \hline

$ \ell $  & 950 & 1750 & 2650 & 3550 & 4450 & 5350 & 6250 & 7150 & 8050\\
\hline
$\rm 250\times 250 $ & $( 8.4\pm 4.8)$$\times$$ 10^{ 13}$  & $( 1.2\pm 0.8)$$\times$$ 10^{ 14}$  & $( 1.6\pm 0.7)$$\times$$ 10^{ 14}$  & $( 1.9\pm 0.9)$$\times$$ 10^{ 14}$  & $( 2.9\pm 0.8)$$\times$$ 10^{ 14}$  & $( 4.2\pm 1.1)$$\times$$ 10^{ 14}$  & $( 5.2\pm 1.1)$$\times$$ 10^{ 14}$  & $( 6.5\pm 1.2)$$\times$$ 10^{ 14}$  & $( 6.6\pm 1.5)$$\times$$ 10^{ 14}$ \\
$\rm 250\times 350 $ & $( 4.2\pm 1.9)$$\times$$ 10^{ 11}$  & $( 8.3\pm 1.9)$$\times$$ 10^{ 11}$  & $( 8.7\pm 2.2)$$\times$$ 10^{ 11}$  & $( 1.3\pm 0.3)$$\times$$ 10^{ 12}$  & $( 1.3\pm 0.2)$$\times$$ 10^{ 12}$  & $( 2.8\pm 0.3)$$\times$$ 10^{ 12}$  & $( 2.8\pm 0.3)$$\times$$ 10^{ 12}$  & $( 3.7\pm 0.4)$$\times$$ 10^{ 12}$  & $( 3.9\pm 0.6)$$\times$$ 10^{ 12}$ \\
$\rm 250\times 500 $ & $( 2.3\pm 0.8)$$\times$$ 10^{ 10}$  & $( 1.9\pm 0.7)$$\times$$ 10^{ 10}$  & $( 2.3\pm 0.6)$$\times$$ 10^{ 10}$  & $( 3.1\pm 1.1)$$\times$$ 10^{ 10}$  & $( 4.5\pm 0.8)$$\times$$ 10^{ 10}$  & $( 5.1\pm 0.8)$$\times$$ 10^{ 10}$  & $( 6.7\pm 1.2)$$\times$$ 10^{ 10}$  & $( 8.4\pm 1.1)$$\times$$ 10^{ 10}$  & $( 8.3\pm 1.5)$$\times$$ 10^{ 10}$ \\
$\rm 250\times 1400 $ &---  & $( 1.9\pm 5.1)$$\times$$ 10^{ 7}$  & $( 1.3\pm 2.1)$$\times$$ 10^{ 7}$  & $( 4.2\pm 5.6)$$\times$$ 10^{ 7}$  & $( 6.7\pm 4.5)$$\times$$ 10^{ 7}$  & $( 9.7\pm 3.8)$$\times$$ 10^{ 7}$  & $( 1.1\pm 0.7)$$\times$$ 10^{ 8}$  & $( 1.2\pm 0.7)$$\times$$ 10^{ 8}$  & $( 1.1\pm 1.0)$$\times$$ 10^{ 8}$ \\
$\rm 250\times 2000 $ & ---  & $( 3.6\pm 2.2)$$\times$$ 10^{ 7}$  & $( 1.1\pm 1.9)$$\times$$ 10^{ 7}$  & $( 1.1\pm 0.9)$$\times$$ 10^{ 7}$  & $( 3.0\pm 1.8)$$\times$$ 10^{ 7}$  & $( 4.6\pm 2.7)$$\times$$ 10^{ 7}$  & $( 2.5\pm 2.8)$$\times$$ 10^{ 7}$  & $( 6.6\pm 4.0)$$\times$$ 10^{ 7}$  & $( 1.4\pm 6.4)$$\times$$ 10^{ 7}$ \\
$\rm 350\times 350 $ & $( 9.1\pm 33.9)$$\times$$ 10^{ 8}$  & $( 6.8\pm 3.4)$$\times$$ 10^{ 9}$  & $( 6.7\pm 2.8)$$\times$$ 10^{ 9}$  & $( 1.0\pm 0.2)$$\times$$ 10^{ 10}$  & $( 1.0\pm 0.5)$$\times$$ 10^{ 10}$  & $( 1.9\pm 0.2)$$\times$$ 10^{ 10}$  & $( 1.8\pm 0.3)$$\times$$ 10^{ 10}$  & $( 3.1\pm 0.4)$$\times$$ 10^{ 10}$  & $( 3.7\pm 0.6)$$\times$$ 10^{ 10}$ \\
$\rm 350\times 500 $ & $( 8.0\pm 6.7)$$\times$$ 10^{ 7}$  & $( 1.7\pm 0.8)$$\times$$ 10^{ 8}$  & $( 1.9\pm 0.5)$$\times$$ 10^{ 8}$  & $( 2.6\pm 0.4)$$\times$$ 10^{ 8}$  & $( 2.8\pm 0.7)$$\times$$ 10^{ 8}$  & $( 3.3\pm 0.5)$$\times$$ 10^{ 8}$  & $( 5.0\pm 0.8)$$\times$$ 10^{ 8}$  & $( 6.8\pm 0.9)$$\times$$ 10^{ 8}$  & $( 6.5\pm 1.6)$$\times$$ 10^{ 8}$ \\
$\rm 350\times 1400 $ & ---  & $( 3.6\pm 3.0)$$\times$$ 10^{ 5}$  & $( 3.5\pm 1.9)$$\times$$ 10^{ 5}$  & $( 4.9\pm 4.0)$$\times$$ 10^{ 5}$  & $( 4.8\pm 2.4)$$\times$$ 10^{ 5}$  & $( 1.0\pm 0.3)$$\times$$ 10^{ 6}$  & $( 1.2\pm 0.4)$$\times$$ 10^{ 6}$  & $( 2.3\pm 0.5)$$\times$$ 10^{ 6}$  & $( 2.0\pm 1.1)$$\times$$ 10^{ 6}$ \\
$\rm 350\times 2000 $ & ---  & $( 2.4\pm 1.6)$$\times$$ 10^{ 5}$  & $( 1.5\pm 1.0)$$\times$$ 10^{ 5}$  & $( 2.1\pm 1.0)$$\times$$ 10^{ 5}$  & $( 3.7\pm 1.3)$$\times$$ 10^{ 5}$  & $( 6.7\pm 1.8)$$\times$$ 10^{ 5}$  & $( 4.6\pm 2.7)$$\times$$ 10^{ 5}$  & $( 9.8\pm 3.9)$$\times$$ 10^{ 5}$  & $( 1.2\pm 0.6)$$\times$$ 10^{ 6}$ \\
$\rm 500\times 500 $ & $( 3.7\pm 0.8)$$\times$$ 10^{ 6}$  & $( 3.9\pm 1.0)$$\times$$ 10^{ 6}$  & $( 5.3\pm 1.0)$$\times$$ 10^{ 6}$  & $( 8.1\pm 1.0)$$\times$$ 10^{ 6}$  & $( 1.0\pm 0.1)$$\times$$ 10^{ 7}$  & $( 1.1\pm 0.2)$$\times$$ 10^{ 7}$  & $( 1.4\pm 0.2)$$\times$$ 10^{ 7}$  & $( 2.3\pm 0.3)$$\times$$ 10^{ 7}$  & $( 2.5\pm 0.5)$$\times$$ 10^{ 7}$ \\
$\rm 500\times 1400 $ &--- & $( 2.1\pm 1.2)$$\times$$ 10^{ 4}$  & $( 1.8\pm 0.7)$$\times$$ 10^{ 4}$  & $( 2.8\pm 1.3)$$\times$$ 10^{ 4}$  & $( 2.9\pm 0.8)$$\times$$ 10^{ 4}$  & $( 4.7\pm 1.0)$$\times$$ 10^{ 4}$  & $( 5.8\pm 1.5)$$\times$$ 10^{ 4}$  & $( 4.9\pm 2.0)$$\times$$ 10^{ 4}$  & $( 6.9\pm 3.1)$$\times$$ 10^{ 4}$ \\
$\rm 500\times 2000 $ & ---  & $( 1.3\pm 0.5)$$\times$$ 10^{ 4}$  & $( 6.7\pm 3.2)$$\times$$ 10^{ 3}$  & $( 1.3\pm 0.4)$$\times$$ 10^{ 4}$  & $( 1.5\pm 0.5)$$\times$$ 10^{ 4}$  & $( 2.7\pm 0.7)$$\times$$ 10^{ 4}$  & $( 3.5\pm 1.1)$$\times$$ 10^{ 4}$  & $( 2.3\pm 1.6)$$\times$$ 10^{ 4}$  & $( 5.9\pm 2.7)$$\times$$ 10^{ 4}$ \\

  \hline
  \end{tabular}
  \caption{Measured $\ell^2 C_{\ell}/2 \pi$ ($\mu \rm K^2$) from Fig.~\ref{fig:bl_spectra} (blue crosses).} 
  \label{tab:dl_data}
\end{sidewaystable*}
%%%%%%%%%% END TABLE 2 %%%%%%%%
%\begin{table*}
%  \centering
%  \begin{tabular}{c||c|c|c|c|c}
%   \hline
% band ($\mu \rm m$) & 250 & 350 & 500 & 1380 & 2030 \\
  % \hline
   %\hline
   % $ 250$ &  1 & $ 7.45\pm 0.XX   $ &  $2.63\pm 0.XX $ &  $ 2.63\pm 0.XX $ & $2.63\pm 0.XX $\\
   % $ 350  $ &  0 & 1 &  $2.63\pm 0.XX $ &  $ 2.63\pm 0.XX  $ & $2.63\pm 0.XX $\\
   % $ 500  $ &  0 & 0 &  1 &  $2.63\pm 0.XX $ & $2.63\pm 0.XX $\\
   % $  1380 $ &  0 & 0 &  0 &  1 & $2.63\pm 0.XX $\\
   % $ 2030 $ &  0 & 0 &  0 &  0 & 1  \\
   % \hline
 % \end{tabular}
 % \caption{Measured Spectral Indices of Poisson and Clustered Components.}
 % \label{tab:si}
%\end{table*}
%%%%%%%%%% END TABLE 2 %%%%%%%%
%\newpage

\begin{acknowledgments}
BLAST was made possible through the support of NASA through grant numbers NAG5-12785,
NAG5-13301, and NNGO-6GI11G, the NSF Office of Polar Programs, the
Canadian
Space Agency, the Natural Sciences and Engineering Research Council
(NSERC) of
Canada, and the UK Science and Technology Facilities Council (STFC).
CBN acknowledges support from the Canadian Institute for Advanced
Research.

ACT was supported by the
U.S. National Science Foundation through awards AST-
0408698 for the ACT project, and PHY-0355328, AST-
0707731 and PIRE-0507768. Funding was also provided
by Princeton University and the University of Pennsylvania.
Computations were performed on the GPC
supercomputer at the SciNet HPC Consortium.  SciNet is funded by: the
Canada Foundation for Innovation under the auspices of Compute Canada;
the Government of Ontario; Ontario Research Fund -- Research
Excellence; and the University of Toronto.
JD acknowledges support
from an RCUK Fellowship. SD, AH, and TM were supported
through NASA grant NNX08AH30G. 
AK was partially supported
through NSF AST-0546035 and AST-0606975 for work on ACT. ES acknowledges support by NSF Physics Frontier
Center grant PHY-0114422 to the Kavli Institute
of Cosmological Physics.
SD acknowledges support from the Berkeley Center for Cosmological Physics. We thank CONICYT for overseeing the Chajnantor Science Preserve, enabling instruments like ACT to operate in Chile; and we thank AstroNorte for operating our scientific base station. Some of the results in this paper have been derived using the HEALPix \citep{gorski/etal:2005} package.

The authors would like to thank Matthieu B\'ethermin and Guilaine Lagache for their help. %  
\end{acknowledgments}
\newpage

%\bibliographystyle{apj}
%\bibliography{refs_1.bib}%apj-jour,refs, act.bib}
%% Here is the endmatter stuff: Supplementary Info, etc.
%% Use \item's to separate, default label is "Acknowledgements"
% ======================

% ======================
%MOVE TO APPENDIX?
\newpage
\appendix
\section{A. Unit Conversion}
\label{sec:convert}
%Define Jansky.  Maps in $\rm Jy~ sr^{-1}$, or $\rm MJy~ sr^{-1}$. \\
The flux density unit of convention for infrared, (sub)millimeter, and radio astronomers is the Jansky, defined as:
\beq
\rm Jy=10^{-26}\frac{\rm W}{\rm m^2~ Hz},
\eeq
and is obtained by integrating over the solid angle of the source.  For extended sources, the surface brightness is described in Jy per unit solid angle, for example, $\rm Jy~ sr^{-1}$, (as adopted by BLAST), or $\rm Jy~ beam^{-1}$ (e.g., SPIRE).  %
%Thus, the flux of objects, $\nuI(\nu)$, is typically given in $\rm Jy$ or $\rm mJy$, while maps pixels are $\rm MJy~ sr^{-1}$.
Additionally, the power spectrum unit in this convention is given in $\rm Jy^2~ sr^{-1}$

CMB unit convention is to report a signal as $\delta T_{\rm CMB}$; the deviation from the primordial $2.73~\rm K$ blackbody.  To convert from $\rm Jy~ sr^{-1}$ to $\delta T_{\rm CMB}$ in $\mu \rm K$, as a function of frequency:
\begin{eqnarray}
\delta T_{\nu} &=& \left(\frac{\delta B_{\nu}}{\delta T}\right), \\
{\rm where}~~ \frac{\delta B_{\nu}}{\delta T} & = & \frac{2k}{c^2}\left( \frac{kT_{\rm CMB}}{h} \right)^2 \frac{x^2e^x}{(e^x -1)^2}= \frac{98.91~ \rm Jy~sr^{-1}}{\mu \rm K} \frac{x^2e^x}{(e^x -1)^2}, \\
{\rm and}~~~~~~~ x &=& \frac{h\nu}{k_{\nu}T_{\rm CMB}}=\frac{\nu}{56.79~\rm GHz},
\end{eqnarray}
\citep{fixsen2009}.  %
Because the BLAST bandpasses have widths of $\sim 30\%$ \citep{pascale2008}, and because the CMB blackbody at these wavelengths is particularly steep (falling exponentially on the Wien side of the 2.73 K blackbody), the integral of $\delta B_{\nu}/\delta T$ over the bands is weighted towards lower frequencies; an effect that becomes dramatically more pronounced at shorter wavelengths.  Thus, the \emph{effective} BLAST band centers in $\delta T$ are $\sim 264, 369$, and 510\rmicron, leading to factors of conversion from nominal of $\sim 2.46$, 1.75, and 1.13, respectively.

%Convert from $\rm Jy~ sr^{-1}$ to $\mu \rm K$ \\
%Convert from $\rm Jy^2$ to $\mu \rm K^2$ \\

Lastly, the CMB power spectrum is conventionally reported versus multipole $\ell$, while in the (sub)millimeter the convention is to report it versus angular wavenumber, $k_{\theta}=1/ \theta$, which is also known as $\sigma$ in the literature, and is typically expressed in $\rm arcmin^{-1}$.  In the small-angle approximation the two are related by $\ell = 2\pi k_{\theta}$.
\section{B. Power spectrum uncertainties}
\label{app:errorbars}
The contents of each map used for cross-frequency correlations can be considered as a sum of two parts: one with a finite cross-correlation; and the other with vanishing cross-power spectrum.
The former contributes to the signal in the cross-power spectrum, while the latter contributes to the uncertainties.
Therefore three terms contribute to the power spectrum uncertainties: sample variance in the signal due to limited sky coverage;
the noise; and a non-Gaussian term due to the Poisson distributed compact sources and galaxy clusters.  The diagonal
component of the  $\rm ACT \times BLAST$ cross-spectrum variance can be written as the sum of these terms, in order:
\be
\label{eq:analyticPSVariance}
\sigma^2( \hat C_b^{\rm A\times B} ) =
          \frac{2}{n_b}\left(\hat C_b^{\rm A\times B}\right)^2 +
          \frac{  \hat C_b (\hat N_b^{(1)} + \hat N_b^{(2)}) + \hat N_b^{(1)} \hat N_b^{(2)} } {n_b}+
    { \frac{ \sigma^2_{\rm P}}{f_{\rm sky}}},
\ee
where $\hat N_b$, estimates the average power spectrum
of the noise; the superscripts ${}^{(1)}$ and ${}^{(2)}$ label the maps (1 for ACT, 2 for BLAST);
$n_b$ counts the number of Fourier modes measured in bin
$b$ (that is, the number of pixels falling in the appropriate annulus
of Fourier space); $f_{\rm sky}$ is the patch area divided by the
full-sky solid angle, $4\pi$\,sr; and $\hat C_b$ is the
mean cross-spectrum.  The last term, $\sigma^2_{\rm P}$, arises from the
Poisson-distributed components in the maps (i.e., unresolved compact
sources and clusters of galaxies) and is
given by the non-Gaussian  part of the four-point
function as described in \citet{fowler2010} and \citet{hajian2010}.  
For purposes of the covariance calculation, we assume that the spatial distribution of these objects is uncorrelated.  
This term is constant with $\ell$.

The noise terms in the ACT and BLAST maps are given by
\bea
\hat N_b^{(1)} &=& C_b^{\rm CMB} +C_b^{\rm RG}+ N_b^{\rm A}, \nonumber \\
\hat N_b^{(2)} &=& N_b^{\rm B},
\eea
where $C_b^{\rm RG}$ is the power spectrum of the radio galaxies in the ACT maps and $N_b^{\rm A}$ and $N_b^{\rm B}$ are the noise spectra in ACT and BLAST respectively.
The noise terms, $N_b$, are dominated by the atmospheric noise on large angular scales and by detector noise on the smallest scales \citep{das2010}. The CMB is a major source of noise for this study out to $\ell \sim 2500$, especially for the \arone{} data.
Radio galaxies only contribute to the uncertainties through the fourth moment of the field. They do not bias the signal. The effect of the radio galaxies is stronger at \arone{} and is negligible at \artwo{}. Therefore we mask the brightest radio galaxies in the \arone{} map to reduce the uncertainty on the cross-power spectra.

The uncertainties on   $\rm BLAST \times BLAST$ power spectra are computed using  a similar analytic estimate, given in Eqn.(9) of \citet{fowler2010} with $n_w=6$ cross-spectra per map. 

As a sanity check, we compare our analytic estimate of the error bars with the standard deviation of the power spectra computed from patches of the sky.  We divide the data into four patches of equal area and with them compute four independent cross-power spectra.  We use the variance of the measurements at each $\ell$ bin as a measure of the error on the power spectrum.  This method agrees well with the analytic estimate of the errors; however, due to the small area of the sky used in this analysis, both analytic and patch-variance estimates of the error bars have uncertainties which are limited by $f_{\rm sky}$.  Thus, we conservatively use the greater of the analytic and patch-variances as an estimate of the uncertainty of the power spectrum.  We test the effect that this choice of error bars has on our results in section \ref{EstimatingBias}.

When fitting parameters, we take the joint likelihood function to be diagonal as the off-diagonal elements are small \citep{das2010}.

% ======================

\end{document}